\documentclass[twocolumn,showpacs,preprintnumbers,amsmath,amssymb,superscriptaddress,aps,prc]{revtex4-1}
\usepackage{graphicx}
\usepackage{xcolor}
\usepackage{subfigure}
\def\scr{\rm\scriptscriptstyle }
\usepackage{amssymb}
\usepackage{graphicx}
\usepackage{subfigure}
\usepackage{amsmath}
\usepackage[mathscr]{eucal}

\def\scr{\rm\scriptscriptstyle }

\begin{document}

\newcommand {\nc} {\newcommand}
\nc {\IR} [1]{\textcolor{red}{#1}}
\nc {\IB} [1]{\textcolor{blue}{#1}}


\title{	Nucleus-nucleus potentials in the scattering of tightly and weakly bound systems}

%
\author{J. Rangel}
\email{jeannierangel@gmail.com}
	\affiliation{Departamento de Matem\'atica, F\'{i}sica e Computa\c{c}\~{a}o Universidade do Estado do Rio de Janeiro, Faculdade de Tecnologia, 27537-000, Resende, Rio de Janeiro, Brazil}
 \author{B. Pinheiro} 
	\email{bpinheiro@id.uff.br}	
	\affiliation{Instituto de F\'{\i}sica, Universidade Federal Fluminense, Av. Litoranea s/n, Gragoat\'{a}, Niter\'{o}i, R.J., 24210-340, Brazil}	
 \author{V. A. B. Zagatto} 
	\email{vzagatto@id.uff.br}	
	\affiliation{Instituto de F\'{\i}sica, Universidade Federal Fluminense, Av. Litoranea s/n, Gragoat\'{a}, Niter\'{o}i, R.J., 24210-340, Brazil}	

\author{J. Lubian} 
\email{jlubian@id.uff.br}
	\affiliation{Instituto de F\'{\i}sica, Universidade Federal Fluminense, Av. Litoranea s/n, Gragoat\'{a}, Niter\'{o}i, R.J., 24210-340, Brazil}
\author{F.M. Nunes} 
\email{nunes@frib.msu.edu}
	\affiliation{Facility for Rare Isotope Beams and Department of Physics and Astronomy, Michigan State University, East Lansing, MI 48824, USA.}

\author{L.F. Canto}
	\email{canto@if.ufrj.br}
	\affiliation{Instituto de F\'{\i}sica, Universidade Federal do Rio de Janeiro, CP 68528, 21941-972, Rio de Janeiro, RJ, Brazil}

\begin{abstract}
\begin{description}
\item[Background] Fusion reactions play an important role in nucleosynthesis and in applications to society. Yet they remain challenging to model.
\item[Purpose] In this work, we investigate the features of the nucleus-nucleus potentials that describe fusion cross sections and compare with those needed for realistic calculations of elastic scattering and other direct-reaction cross sections. 
\item[Method] We perform coupled-channel calculations for studying elastic and fusion reactions around the Coulomb barrier with a tightly bound projectile ($^{16}$O+$^{144}$Sm). We also perform  Continuum Discretized Coupled Channel calculations to study elastic ($^{8}$B+$^{58}$Ni) and fusion ($^{6}$Li+$^{198}$Pt) of loosely bound projectiles  in the same energy regime.
\item[Results] We constrast the coupled-channel results with those obtained in a single-channel solution with different assumptions for polarization potentials to shed light on the relevant absorption terms required for the two different reaction channels.
\item[Conclusions] Our results suggest that different approximations may be required for modeling direct processes and for modeling fusion reactions.
\end{description}
\end{abstract}

\maketitle


\section{Introduction}

Fusion reactions are of great interest because of their
important role in the evolution of the elements. They are also relevant in applications of nuclear physics to
society. From a theoretical perspective, fusion reactions continue to pose conceptual and implementation
challenges to the theory community. Sometimes, it is
necessary to connect back to simple models that help
discriminate what the essential elements are to describe
fusion, and how they differ from the assumptions of
direct reaction channels. \\

Over the last four decades, numerous studies have described the collisions of tightly bound projectiles with heavy targets, with a focus on fusion reactions~\cite{Hag20}. It is well understood that projectile and target excitations are important pathways for fusion, especially for beam energies around the Coulomb barrier. 
Furthermore, collisions of light weakly bound projectiles with heavy targets have also attracted great interest in the last decades~\cite{TBD85,CGD06,KRA07,KAK09,ThN09,CGD15,KGA16,HOM22}. 
However, there are two important differences between weakly and tightly bound systems. The first is that the low breakup threshold of the projectile leads to a lower Coulomb barrier. 
The second is that the reaction dynamics involving weakly bound nuclei is strongly influenced by the breakup process. \\

Several experimental and theoretical studies of elastic scattering, 
breakup, and fusion of weakly bound systems have been conducted (e.g.~\cite{CGD15,KGA16,HOM22}). In this work, we will focus only on projectiles that break up into two fragments, such as $^{6,7}$Li, although similar arguments could be made for more complex structures. The important point to understand is that weakly bound systems offer additional pathways to fusion that tightly bound systems do not. In addition to the usual complete fusion reaction (CF), observed for tightly bound systems, there are incomplete fusion (ICF) reactions, following the breakup of the projectile, where one of the projectile's fragments fuses with the
target, while the other one does not. Determining these various fusion cross sections remains a great challenge for experimentalists and theorists.\\

From a theoretical perspective, one must evaluate the breakup cross section and its impact on fusion, elastic scattering, and other direct reactions. For this purpose, the most successful tool is the coupled-channel (CC) method, where one adopts a model Hamiltonian involving just the projectile-target separation vector and a few intrinsic coordinates of the projectile. The total wave function is then expanded over a basis of intrinsic states, and the full Schr\"odinger equation is reduced to a set of coupled equations, which can be solved
numerically (the two codes widely used are readily available~\cite{Tho88,HRK99}). However, when breakup channels are included in the channel expansion, one has to deal with the continuum spectrum of the projectile, and the CC method leads to an infinite set of coupled
equations. One way to overcome this difficulty is by discretizing the continuum (e.g., within the continuum discretized coupled channel (CDCC) method~\cite{SYK83,SYK86,AIK87}).\\

It is noteworthy that neither the CC nor the CDCC expansions can explicitly describe the fusion channel. There is no way to include explicitly the large number 
of intermediate states that are part of the pathways from the entrance channel to a fused and equilibrated compound
nucleus (CN). The coupled-channel framework computes fusion by associating it with the part of the incident flux
that gets absorbed, \textit{i.e.}, the part that does not emerge from the interaction region following the reaction.
In practice, the effect of fusion on elastic scattering and direct reactions can be
simulated by adopting incoming-wave boundary conditions at a radius in the inner region 
of the Coulomb barrier~\cite{Raw64}, or by adding a strong short-range imaginary potential to the 
Hamiltonian. This short-range imaginary potential simulates the region where the projectile-target nuclear densities overlap and where the model Hamiltonian breaks down. This imaginary potential should be negligible elsewhere. \\

Suppose that one adopts a realistic projectile-target interaction and includes all relevant channels in the expansion. In that case, the CC method is expected to lead to a realistic description of all reaction data, including fusion. However, in typical CC calculations, the channel expansion is truncated. Furthermore, only the main collective channels are considered.
In most situations, it leaves out a large number of inelastic and/or transfer channels that are weakly
coupled to the entrance channel. Although they may be irrelevant individually, their combined influence on the collision dynamics may be important \citep{CGZ22}. The situation becomes even more complicated in collisions involving weakly bound projectiles. In this case, the bound and unbound excited states of the projectile can be taken into account in CDCC calculations. 
%
 However, although there are CDCC implementations that allow excitations of the target and the cores 
(e.g.~\cite{GoM17,SNT14}), the computational cost of including projectile and target excitation in CDCC would be prohibitive. 
For realistic calculations of elastic scattering and nuclear reaction cross sections, it is necessary to account for the influence of these channels, at least within some approximation. The question we address in this work is whether a suitable approximation for elastic scattering and direct reactions remains appropriate for calculating fusion cross sections. \\

This paper is organized as follows. In Sect. II, we review some basic concepts of heavy-ion scattering. We briefly discuss the main ingredients of potential scattering and the CC method for tightly bound and weakly bound systems. We then introduce the concept of polarization potentials and discuss their application in practical studies of heavy-ion collisions.  In Sect. III, we discuss imaginary potentials suitable for theoretical descriptions of elastic scattering of tightly and weakly bound systems 
at near-barrier energies. In Sect. IV, we conduct a similar discussion for the calculations of fusion cross sections. Finally, in Sect. V, we summarize the 
conclusions of this work.



\section{Theory of elastic scattering and direct reactions in heavy-ion collisions}\label{Theory}

 
 In distant and peripheral collisions at near-barrier energies, the many-body Hamiltonian can be replaced by a model Hamiltonian involving 
 only the projectile-target separation vector, ${\bf R}$, as a dynamical variable, and a few intrinsic degrees of freedom, $\xi$. This Hamiltonian can be written as
 \begin{equation}
 \mathbb{H} = T+h+\mathbb{U},
 \label{Hmodel}
 \end{equation}
where $T=-\, \hbar^2\,\nabla^2/2\mu$ is the kinetic energy operator associated with ${\bf R}$, with $\mu$ standing for the reduced mass of the system, and $h\equiv h(\xi)$ is the intrinsic Hamiltonian. In the above equation, $\mathbb{U} \equiv \mathbb{U}({\bf R},\xi)$ denotes the  total interaction 
 between the projectile and the target.
The total wave function of the system, $\Psi ({\bf R},\xi)$, satisfies the Schr\"odinger equation\\
 \begin{equation}
 \Big[ E-\mathbb{H}({\bf R},\xi) \Big]\, \Psi ({\bf R},\xi)= 0.
 \label{Sch1}
 \end{equation}

In the CC method, one carries out the channel expansion
\begin{equation}
{\rm \Psi} ({\bf R},\xi) = \sum_{\alpha=0}^{N-1}  \psi_\alpha({\bf R})\times \phi_\alpha (\xi).
\label{ch-exp}
\end{equation}
Above, $\psi_\alpha({\bf R})$ is the wave functions that describe the relative motion of the projectile-target system in channel $\alpha$, and $\phi_\alpha (\xi)$ is the corresponding eigenstate 
of the intrinsic Hamiltonian. Using Dirac notation ($\phi_\alpha (\xi)\,\rightarrow |\alpha )$), the intrinsic states satisfy the equations
\begin{equation}
h\, \left| \alpha \right) = \varepsilon_\alpha\,\left| \alpha \right)\qquad \left( \alpha \vert  \alpha^\prime\right) = \delta_{\alpha,\alpha^\prime}.
\end{equation}

In actual calculations, the channel expansion is truncated after a finite number of channels. In Eq.~(\ref{ch-exp}) the expansion includes $N$ channels, labeled
$\alpha = 0,1, ...,N-1$, with $\alpha = 0$ representing the elastic channel. 
Inserting this expansion in Eq.~(\ref{Sch1}), taking the scalar product with each intrinsic state and using the orthogonality property, one gets the set of CC equations,
\begin{equation}
\left[  E_{\alpha} -T -U_{\alpha} (\bf{R)}\right]  \,\psi_{\alpha}(\mathbf{R})= 
\sum_{\alpha^\prime \ne \alpha=0}^{N-1}\,U_{\alpha\alpha^\prime}(\bf{R)}\ \psi_{\alpha^\prime}(\mathbf{R}) ,
\label{CC-eqs}
\end{equation}
where
\begin{equation}
U_{\alpha\alpha^\prime}(\bf{R)} = \int d\xi\ \phi_\alpha^*(\xi)\ \mathbb{U}({\bf R},\xi)\  \phi_{\alpha^\prime}(\xi)
\end{equation}
and
\begin{equation}
E_\alpha = E - \varepsilon_\alpha.
\end{equation}
In Eq.~(\ref{CC-eqs}), we have used a shorthand notation $U_{\alpha}(\mathbf{R}) \equiv U_{\alpha\alpha}(\mathbf{R})$. A necessary condition for 
accurately describing the collision is that the channel expansion includes all relevant direct reaction channels.\\

The channel wave functions, $\psi_\alpha$, are solutions of the CC equations that obey scattering boundary conditions. 
They have an asymptotic behavior 
\begin{equation}
\psi_\alpha\left(\mathbf{R}\right)  \rightarrow N
\left[
\phi_{\scr C}\left(\mathbf{R}\right)  \,\delta_{\alpha 0%
}\,+\,\bar{f}_{\alpha,0}(\mathbf{k}_{\alpha},\mathbf{k}%
_{0})\ \frac{e^{i\Theta_{\alpha}\left(  R\right)  \,}}{R}\right].
\label{8cc-psias-c}%
\end{equation}
Above, $N$ is an arbitrary normalization constant, $\phi_{\scriptscriptstyle\rm C}$ is the wave function for Coulomb scattering,
$\bar{f}_{\alpha,0}(\mathbf{k}_{\alpha},\mathbf{k}_{0})$ is the partial scattering amplitude arising from the short-range potential,
and $\Theta_{\alpha}\left(  r\right)$ is the asymptotic phase of the outgoing wave. It is given by
\begin{equation}
\Theta_{\alpha}\left(  r\right)  =k_{\alpha}r-\eta_{\alpha}\ln(2k_{\alpha}r),
\label{8Theta-alpha}%
\end{equation}
where $k_\alpha$ and $\eta_{\alpha}$ are the wave number and the Sommerfeld parameter in channel-$\alpha$, respectively (see
Refs.~\cite{CaH13,ThN09} for details).\\

%
%

For practical purposes, the total wave function is expanded in angular momentum, and the radial wave functions, $u_{\alpha l}\left(k_\alpha R\right)$  (for simplicity, we neglect spins at this stage) are determined numerically. From their asymptotic form, one gets the components of the nuclear S-matrix, $\overline{S}_{\alpha,\alpha'}(l)$, which are used to determine the 
cross sections for the channels included in the expansion of Eq.~(\ref{ch-exp}). In particular, the elastic scattering cross section is given by the expression
\begin{equation}
\frac{d\sigma _{\rm el}(\Omega )}{d\Omega }=\left| f_{\scr C}(\theta )+\bar{f}_{0}(\theta
)\right| ^{2},  
\label{cross}
\end{equation}
where $f_{\scr C}(\theta )$ is the well-known Coulomb amplitude and $\bar{f}_{ 0}(\theta )$ is the short-range part of the elastic scattering amplitude. It is given by the partial-wave expansion
\begin{equation}
\bar{f}_0(\theta ) = \frac{1}{2ik}\,\sum_{l=0}^{\infty }\,(2l+1)\,\,P_{l}(\cos \theta
)\,\,e^{2i\sigma _{l}}\,\left[ \bar{S}_0(l)-1\right] , 
\label{fbar2}
\end{equation}
where $\sigma_{l}$ is the Coulomb phase-shift at angular momentum $l$, and $k$ is the wave number corresponding to the collision energy, $E$. In Eq.~(\ref{fbar2}), we used a shorthand notation $\bar{S}_{0,0}(l) = \bar{S}_0(l)$. Note that this cross section is entirely determined by the asymptotic behavior of the elastic wave function.\\

\subsection{Bound and continuum states}
\label{sec:breakup}


Collisions of weakly bound projectiles are more challenging to describe. The low breakup threshold results in large breakup cross sections, and the breakup channels exert a significant influence on the collision dynamics. Then, one has to deal with unbound intrinsic wave functions that have infinite norms and are labeled by a continuous excitation energy. However, these difficulties are overcome in the CDCC method~\cite{SYK83,SYK86,AIK87}. It consists of approximating the continuum states of the projectile by a finite set of wave packets, usually called {\it bins}. 
In this way, one obtains a finite set of coupled-channel equations that can be handled using the same procedures followed in the standard coupled-channel problem.\\

In the CDCC method, the channel space can be split into a bound (B) and a continuum-discretized sub-space (C). Then, the total wave  function can be written as,
\begin{equation}
{\rm \Psi} = {\rm \Psi}^{\rm B}\ + \   {\rm \Psi}^{\rm C},
\label{Psi-B_Psi-C}
\end{equation}
where $ {\rm \Psi}^{\rm B}$ and $ {\rm \Psi}^{\rm C}$ are respectively the components of the wave function in the bound and bin subspaces. They are given by the expansions
\begin{equation}
{\rm \Psi}^{\rm B} =  \sum_{\alpha \in B}  \psi_\alpha({\bf R}) \left|\alpha\right); \qquad 
{\rm \Psi}^{\rm C} =  \sum_{\beta \in C}  \psi_\beta ({\bf R}) \left|\beta\right).
\label{expB-expC}
\end{equation}
While ${\rm \Psi}^{\rm B}$ is used to evaluate the elastic cross section and the cross sections for the other bound channels 
in the expansion, ${\rm \Psi}^{\rm C}$ is used to determine the breakup cross section.\\

We emphasize that the validity of the CC approach is based on two assumptions. The first is that the model Hamiltonian of 
Eq.~(\ref{Hmodel}) gives a realistic description of the collision. This assumption is reasonable when the distance between the 
projectile and the target remains greater than the radius of the Coulomb barrier. However, it breaks down when the nuclear densities
of the collision partners overlap strongly. When this happens, there is a large exchange of nucleons between the projectile and the 
target, leading to final configurations that cannot be described by the model Hamiltonian of Eq.(\ref{Hmodel}). The main outcome of 
such collisions is a fusion reaction. To describe fusion, one typically adds a strong short-range imaginary potential to the projectile-target interaction.\\

The second condition is that the channel expansion must include all channels that
affect the reaction dynamics. In typical heavy-ion collisions, this condition is not
fully satisfied. Although the main couplings are taken into account, many weakly
coupled channels are neglected. One difficulty in including them all in the calculation is
that their coupling matrix elements are not well known. Then, the standard procedure is to use optical potentials that include an imaginary term to effectively account for coupling to those channels that are not explicitly included in the model space. This issue will be addressed in the forthcoming sections of this paper. \\

\subsection{Fusion absorption}
\label{sec:fusabs}


The fusion process plays an important role in heavy-ion collisions. The effects of the fusion channel on elastic 
scattering and direct reactions may be simulated by the addition of a short-range imaginary potential, 
$\mathbb{W}_{\rm F}$, to the model Hamiltonian.  It is usually assumed to be diagonal in channel space, independent 
of $\alpha$, and spherically symmetric. One writes,\\
\begin{equation}
U_\alpha(R) = V_\alpha(R) - i\,W_{\rm F}(R).
\label{W_F}
\end{equation}
$W_{\rm F}(R)$ must be very strong to produce total absorption in the inner region of the Coulomb barrier, but it should
be negligible elsewhere. 
Usually, one adopts a short-range Woods-Saxon (WS) function of the form
 \begin{equation}
 W_{\rm F}(R) = \frac{W_0}{1+ \exp\left[\left(R-R_{\rm w}\right)/a_{\rm w} \right]},
  \label{WF1}
 \end{equation}
 with 
 \begin{equation}
 R_{\rm w} = r_{\rm w}\ \left[ A_{\rm P}^{1/3} + A_{\rm T}^{1/3}  \right],
 \label{WF2}
 \end{equation}
  where $A_{\rm P}$ and $A_{\rm P}$ stand for the mass number of the projectile and target, respectively. Typical values of the WS parameters (depth, reduced radius, and diffuseness) for strong short-range absorption are
 \begin{equation}
 W_{0} = 50\,{\rm MeV}, r_{\rm w} = 1.0\,{\rm fm,\ \ and}\ a_{\rm w} = 0.2\,{\rm fm}.
 \label{WSpar}
 \end{equation}
However, the values of these parameters may be slightly different from those listed above, provided that the imaginary
potential simulates fusion absorption~\cite{CDH18}.
The results should not depend much on the details of this absorption term. Its primary function is to impose the incoming boundary condition on the solution. The assumption is that once 
the system penetrates the barrier, it gets absorbed. For the validity of this assumption, it is necessary that the collision energy does not exceed a critical value, above which there is
no longer a grazing angular momentum. That is, the potential $V_{l \,>\, l_g}(R)$ ceases to have a pocket. Further, it is not valid for very heavy systems, where fusion may require an extra push above the Coulomb barrier~\cite{ZAL25,Swi82}. These conditions are satisfied in the collisions studied in the present work.

\subsection{The classical fusion cross section}
\label{sec:classicalfusion}

The simplest quantum-mechanical description of heavy-ion collisions is that of single-channel scattering. It corresponds 
to neglecting all channel couplings, so that the CC equations are reduced to a single equation for the elastic channel. 
In this case, assuming that the potential is spherically symmetric, the elastic wave function satisfies the one-channel
equation,\\
\begin{equation}
\Big[ T + V(R) - i\,W_{\rm F}(R) \Big]\, \psi({\bf R}) = E \, \psi({\bf R}).
 \end{equation}

Carrying out an angular momentum expansion, one gets an equation for the radial part of 
the wave function (see, e.g., Ref.~\cite{CaH13}). The asymptotic form of this wave 
function,  $u_l(k,R)$, gives the corresponding component of the nuclear S-matrix, 
$\overline{S}_l$, in terms of which one obtains the elastic and the fusion cross sections.
The elastic angular distribution is then given by Eqs.~(\ref{cross}) and ({\ref{fbar2}}), 
with the replacements: \\
\[
\overline{f}_0(\theta)\ \longrightarrow\ \overline{f}(\theta)\ \ {\rm and}\ \  
\overline{S}_0(l)\ \longrightarrow\  \overline{S}_l.
\]

\bigskip

In this approach, fusion corresponds to absorption by the imaginary potential 
$W_{\rm F}(R)$. Thus, the fusion cross section is given by
\begin{equation}
\sigma_{\rm F} =  \frac{\pi}{k^ 2}\,\sum_{l=0}^{\infty }\,(2l+1)\, P_{\rm F}(l),  
\label{sigF0}
\end{equation}
where 
\begin{equation}
P_{\rm F}(l) = 1\, -\, \left| \bar{S}(l) \right|^2  
\label{Pf1}
\end{equation}
is the fusion probability in a collision with energy $E$ and angular momentum $l$. This probability can also be evaluated by the integral of the radial wave function~\cite{CaH13},\\
\begin{equation}
P_{\rm F}(l) =  \frac{4k}{E} \int dr\ \left| u_{l}(k,R) \right|^2\ W_{\rm F}(R) .  
\label{sigF2}
\end{equation}
where $W_F$ is the potential introduced in Eq. (\ref{W_F}).

The fusion cross section calculated in this way is practically identical to the one obtained by the barrier penetration model (BPM)~\cite{CDH18}. This model assumes that the incident current 
that reaches the inner region of the barrier is fully absorbed, leading to fusion. Then, the fusion probability can be approximated by the
transmission coefficient through the barrier of the total real potential,
\begin{equation}
V_l(R) = V(R)+\frac{\hbar^2}{2\mu\,R^2}\ l(l+1).
\label{Vl}
\end{equation}
That is,
\begin{equation}
P_{\rm F}(l) \simeq T_{\rm F}(l).
\label{PF-TF}
\end{equation}
The transmission coefficient is frequently evaluated using the improved Kemble WKB method~\cite{Kem35,CGD06}.\\

So far, we have neglected the spins of the collision partners. In realistic calculations, it may be necessary to take them into account. Then, one gets similar expressions for the cross sections, but involving the total angular momentum quantum number, $J$, and some angular momentum coupling coefficients (see, e.g., Refs.~\cite{TNL89,ThN09,CaH13}). \\

The fusion cross section takes a straightforward form if one neglects tunnelling effects, adopting the classical
approximation for the transmission coefficient. Then, the results become:
\begin{equation}
P_{\rm F}(l) = 1, \ {\rm for}\ l \le l_g ;\qquad P_{\rm F}(l) = 0, \ {\rm for}\ l > l_g,
\label{PFclass}
\end{equation}
where, $l_g$ is the grazing angular momentum, defined as the $l-$value for which the barrier of $V_l(R)$, $B_l$,
is equal to the collision energy. This approximation is reasonable, except at sub-barrier energies, or 
energies just above the barrier. Its region of validity can be expressed as: $E \gtrsim V_{\rm B} + 2\,{\rm MeV}$~\cite{CZL23},
where $V_{\rm B}$ is the height of the Coulomb barrier. 
With this approximation, the partial-wave series becomes an arithmetic progression, which can be 
summed analytically. Then, using the approximation, $(l_g+1)^2\simeq \l_g^2$, which is valid in this
energy range, we get~\cite{CZL23}\\
\begin{equation}
\sigma_{\rm cl} = \pi\ \frac{l_g^2}{k^2}.
\label{sigF-lg}
\end{equation}
%


\subsection{Polarization potentials} \label{sec:polarization}


The number of CC equations can be reduced by introducing polarization potentials. For this purpose, the channel space
is divided into two subspaces, $\mathcal{P}$ and $\mathcal{Q}$. Usually, the first, with dimension $n$, contains the
channels that are strongly coupled to the entrance channel. In contrast, the second, with dimension $m=N-n$, contains the
remaining channels that may still affect the reaction dynamics but cannot be explicitly included. These subspaces are associated with the projectors~\cite{Fes62}
\begin{equation}
P = \sum_{\alpha = 0}^{n-1} \left| \alpha \right)\,\left( \alpha \right|;\ \ \ Q = \sum_{\alpha = n}^{N} \left|\alpha \right)\,\left( \alpha \right|,
\label{P-Q_def}
\end{equation}
that have the usual properties,
\begin{equation}
P^2 = P;\ \ Q^2=Q, \ \ P Q = Q P =0\ \ {\rm and} \ \ P+Q = I, 
\end{equation}
where $I$ is the identity operator. The scattering wave function can be split into its components in the two subspaces.
\begin{equation}
\left| {\rm \Psi} \right> \equiv \left(P+Q\right)\,\left| {\rm \Psi} \right> = \left| {\rm \Psi}_P \right> + \left| {\rm \Psi}_Q \right>.
\label{PsiP-PsiQ}
\end{equation}
where\\
\begin{equation}
\left| {\rm \Psi}_P \right> \equiv P\left| {\rm \Psi} \right>,\qquad \left| {\rm \Psi}_Q \right> \equiv Q\left| {\rm \Psi} \right>.
\end{equation}

Using this decomposition, the full CC equations can be reduced to an equation in the $\mathcal{P}$-space, with
the effective Hamiltonian~\cite{Fes62}
\begin{equation}
\mathbb{H}_{\rm eff}  = T\,+\,\mathbb{U}_{\scr PP}\,+\,\mathbb{U}_{\scr pol},
\label{HHeff}
 \end{equation}
with the polarization potential
\begin{equation}
\mathbb{U}_{\scr pol} = \mathbb{U}_{\scr PQ}\,G^{\scr (+)}_{\scr QQ}\, \mathbb{U}_{\scr QP}\equiv \mathbb{V}_{\scr pol} - i\, \mathbb{W}_{\scr pol} ,
 \end{equation}
Above, the subscripts $P$ and $Q$ indicate the action of the projection operators from the left and from the 
right, and $G^{\scr (+)}_{\scr QQ}$ is the Green's function in the $\mathcal{Q}$-space.\\

The imaginary part of the polarization potential, $\mathbb{W}_{\scr pol}$, which reaches the barrier region and
beyond, accounts for attenuation of the incident flux, resulting from transitions to channels in the 
$\mathcal{Q}$-subspace. After the transitions, the incident current in these channels may be reflected at 
the barrier or reach the inner region. Thus, they contribute to the direct reaction and also to the fusion 
cross sections. However, the polarization potential approach cannot separate these two components.
For this reason, this approach may not lead to reliable calculations of $\sigma_{\rm F}$. This issue will be 
investigated in Sect.~\ref{Fusion}. \\

An exact evaluation of $\mathbb{U}_{\scr pol}$ would be too complicated. It is non-local and strongly 
state-dependent. However, several local approximations have been proposed~\cite{HBC84,FrE87,CCD21}. 
The realistic polarization potential of Thompson, Nagarajan and Lilley~\cite{TNL89} is discussed below.\\

Polarization potentials are simpler when $P$ projects exclusively on the elastic channel. In this case, 
Eq.~(\ref{P-Q_def}) becomes
\begin{equation}
P = \left| 0 \right)\,\left( 0 \right|;\ \ \ Q = \sum_{\alpha = 1}^{N-1} \left|\alpha \right)\,\left( \alpha \right|.
\label{P-Q_1}
\end{equation}
Then, taking the expectation value of the Hamiltonian of Eq.~(\ref{HHeff}) with respect to the elastic channel and adopting the coordinate representation, one gets the following.
\begin{equation}
H  = T\,+\,U(R)\,+\, U_{\scr pol}({\bf R},{\rm \bf R}^\prime),
\label{Heff-1}
\end{equation}
where
\begin{equation}
U(R) = \left( 0 \right|\,\mathbb{U}({\bf R},\xi)\,\left| 0 \right),
\label{Ubare}
\end{equation}
and \\
\begin{equation}
U_{\rm pol}({\bf R},{\bf R}^\prime) = \sum_{\alpha,\alpha^\prime=1}^{N-1} U_{0\alpha}({\bf R})\
G^{(+)}_{\alpha,\alpha^\prime}({\bf R},{\bf R}^\prime)\ U_{\alpha^\prime 0}({\bf R}^\prime).
\label{UpolRRprime}
\end{equation}

For practical purposes, one performs angular-momentum projections of the wave function and polarization potential. Taking spins into account, the elastic wave function and the polarization potential 
are expressed in terms of their components, $u_{lJ\pi}(R)$ and $U_{\rm pol}^{l l^\prime J\pi}(R,R^\prime)$, 
respectively. \\

The polarization potential of Eq.~(\ref{UpolRRprime}) has the drawback of being non-local. We can introduce
a trivially equivalent local potential, $\overline{U}_{\rm pol}^{\ l l^\prime J\pi}(R)$,
defined by the condition,
\begin{equation}
\overline{U}_{\rm pol}^{\,l l^\prime J\pi}(R) = \frac{1}{u_{lJ\pi}(R)}
\int dR^\prime\ U_{\rm pol}^{l l^\prime J\pi}(R,R^\prime)\ u_{lJ\pi}(R^\prime).
\label{loca_U}
\end{equation}
However, this potential has serious shortcomings: first, it is angular-momentum dependent, and second, it exhibits poles at the zeros of the radial wave function.\\

These problems were eliminated in the polarization potential proposed by Thompson, Nagarajan, and Liley~\cite{TNL89}.
It is the $l$-independent \lq weighted mean\rq\ local polarization potential given by
\begin{equation}
\overline{U}_{\rm pol}(R) = \frac{
\sum_{ll^{\prime}J\pi}\ w_{ll^\prime J\pi}(R)\ \overline{U}_{\rm pol}^{\,ll^\prime J\pi}(R)}
{\sum_{ll^\prime J\pi}\ w_{ll^\prime J\pi}(R)}
,  
\label{Vpol-Th}
\end{equation}
where $\overline{U}_{\scr pol}^{\,ll^\prime J\pi}(R)$ are the trivially equivalent local potentials of Eq.~(\ref{loca_U}), and the 
$w_{ll^\prime J\pi}(R)$ are the weight factors,
\begin{multline}
w_{ll^\prime J\pi}(R)=\sqrt{(2l+1)(2l^\prime +1)}\\
\times \Big[\, 1-\big| S_{ll^\prime J\pi}\big|^{2}\,  \Big]\ u^*_{l^\prime J\pi}(R) \
u_{lJ\pi}(R).
\label{weight}
\end{multline}
Note that the weight factors vanish at the zeros of the radial wave function. In this way, the divergences
of the trivially local equivalent potentials are removed. The calculation of this polarization potential is
implemented in the {\sc fresco} code~\cite{Tho88}  and will be adopted in the remaining part of this paper. 
It has been applied to several collisions~\cite{TNL89,RLC16,ZLG17}.\\

The polarization potential approach leads to the one-channel equation,
\begin{equation}
\Big[ T + U_{\rm eff}(R) \Big]\, \psi({\bf R}) = E \, \psi({\bf R}),
\label{eff-1ch}
 \end{equation}
where 
\begin{equation}
U_{\rm eff}(R) = U \,+\,U_{\rm pol},
\label{Ueff}
\end{equation}
with\\
\begin{eqnarray}
U(R) &=& V(R)\,- i\, W_{\rm F}(R) \label{U},\\
U_{\rm pol}(R) &=& V_{\rm pol}(R) \,-\,i\,W_{\rm pol}(R)
\label{Upol}.
\end{eqnarray}
The {\it  bare} potential, $V(R)$, can be evaluated by systematic procedures, like the folding~\cite{SaL79} and proximity~\cite{BRS77} models. Since they are determined
by the matter distributions of the colliding nuclei, they are, in principle, energy-independent. The S\~ao Paulo~\cite{CPH97,CCG02} and the Aky\"uz-Winther~\cite{BrW04,AkW81}  potentials are 
frequently used implementations of the folding model (note that the SPP has a weak energy-dependence that arises from the Pauli Principle). Making several approximations on the folding integral, 
Aky\"uz and Winther obtained a potential given by a simple expression, in terms of parameters depending on the atomic and mass numbers of the collision partners. On the other hand, the SPP implementation uses realistic densities of the collision partners and evaluates the folding integrals numerically.  In this way, the AW potential is less accurate but easier to use, whereas the SPP is more realistic. However, its evaluation requires a specific implementation and is not embedded in general-purpose codes. The imaginary part of $U(R)$, $W_{\rm F}(R)$,  is usually represented by a short-range WS function, with the 
parameters of Eq.~(\ref{WSpar}). \\

On the other hand, $U_{\rm eff}(R)$ contains channel-coupling effects in its polarization potential
component. Therefore, it depends on the system and the collision energy. Some systematic studies of elastic 
scattering lead to average energy-dependent parameterizations for $U_{\rm eff}(R)$ that satisfactorily reproduce
the elastic scattering data. Several works along these lines can be found in the literature~\cite{ZLG17,ZGG22}. Typically, these potentials are
designed for a specific projectile in a given energy range. Koning and Delaroche~\citep{KoD03} developed a 
phenomenological optical model potential for proton and neutron scattering, expressed in terms of several 
energy-dependent parameters. Perey and Perey~\citep{PeP76} have also made a compilation of optical potentials 
for several projectiles, like protons, neutrons, deuterons, tritons, and $^{3,4}$He. Optical potentials for heavier systems were included in the systematic work of Alvarez {\it et al.} \cite{ACH03}. These authors adopted a parametrization in the form $U(R) = \left[N_{r} +i\, N_i \right] \times V_{\rm SPP}(R)$, where
$V_{\rm SPP}$ stands for the S\~ao Paulo potential. The above-mentioned potentials were widely used to describe
cluster-target potentials in CDCC calculations of elastic scattering, inelastic scattering and breakup, in
collisions of weakly bound projectiles~\cite{OLG09,GoM17,ZLG17,PSS17,ARG18,ZGG22}.\\

The polarization potential of Thompson, Nagarajan, and Liley leads to a good approximation to the exact elastic wave function. Although it may appear impractical, since the full CC wave function is required to evaluate the polarization potential itself (see Eqs.~(\ref{Vpol-Th}) and (\ref{weight})), it can be beneficial to reduce the dimension of the channel space in more complicated CC calculations, as discussed below.\\ 

Let us consider a CC problem with the projectors of Eq.~(\ref{P-Q_1}), where the $\mathcal{P}$-subspace, with dimension $n$, is split into two pieces. The first, denoted by $\mathcal{P}_0$, contains only the elastic channel.
The second, $\mathcal{P}_1$, contains the remaining $n-1$ channels of $\mathcal{P}$ ($\alpha =1, ...,n-1$).
The projectors involved in the reaction dynamics are then given by, 
\begin{equation}
P_0 = \left| 0 \right)\,\left( 0 \right|;\ \ \ P_1 = \sum_{\alpha = 1}^{n-1} \left|\alpha \right)\,\left( \alpha \right|;
\ \ \ Q = \sum_{\beta = n}^{N} \left|\beta \right)\,\left( \beta \right|.
\label{P-Q_1-Q_2}
\end{equation}
We can solve the CC problem in a two-step approximation. First, a polarization potential is evaluated that accounts for the influence of 
the $\mathcal{Q}$-space on the elastic channel, $\overline{U}_{\rm pol}^{\,\rm Q}(R)$. For this purpose, we evaluate the non-local potentials
by Eqs.~(\ref{UpolRRprime}) and (\ref{loca_U}), with $\alpha$ running over the channels in the 
$\mathcal{Q}$-space, and then use them in Eq.~(\ref{Vpol-Th}). 
In the second step, we solve the CC problem in the $\mathcal{P}$ sub-space, including this polarization potential in the diagonal part of the 
Hamiltonian. The CC equations then read\\
\begin{multline}
\left[  E_{\alpha}+T -U_{\alpha} ({\bf R)} - \overline{\it U}_{\rm pol}^{\,\rm Q}(R) \right]  \,\psi_{\alpha}(\mathbf{R}) \\
= 
\sum_{\alpha^\prime \ne \alpha=0}^{n-1}\,U_{\alpha\alpha^\prime}(\bf{R)}\ \psi_{\alpha^\prime}(\mathbf{R}) .
\label{CC-eqs-1}
\end{multline}
Note that this procedure does not provide the exact solution to the full CC problem, since couplings among the channels in the $\mathcal{P}_1$ and the $\mathcal{Q}$ subspaces are not directly taken into account.  Nevertheless, we found it to be effective in describing reactions~\cite{ZLG17,ZGG22}.\\

The procedure discussed above is particularly useful in calculations of elastic scattering and breakup cross sections 
in collisions of weakly bound projectiles with heavy targets, where the breakup channels play a major role in the 
reaction dynamics. In such cases, one should perform CDCC calculations, considering also excitations of the target. 
However, as we mentioned in the introduction, the computational cost of such a calculation would be forbidding.
Then, to remedy the situation, one can replace couplings with excitations of the target by a polarization potential. 
That is, one first evaluates the polarization potential that accounts for the influence of the target's excitations on 
the elastic channel, in the absence of breakup. Then, we include this potential in the diagonal part of the Hamiltonian 
of the CDCC equations. Applications of this procedure for elastic scattering and direct reactions have been
performed, either using calculated polarization potentials~\cite{ZLG17} or polarization potentials fitted to data (see e.g., Refs.~\cite{PLO09,CCB13}).\\


\section{Elastic scattering and direct reactions}


Now we discuss the real and imaginary parts of the projectile-target interaction in elastic scattering 
calculations. Adopting the polarization potential approach, the scattering wave function
satisfies Eq.~(\ref{eff-1ch}), with the effective potential of Eqs.~(\ref{Ueff}), (\ref{U}), and (\ref{Upol}).\\

\begin{figure}
\begin{center}
\includegraphics[width=8.5 cm]{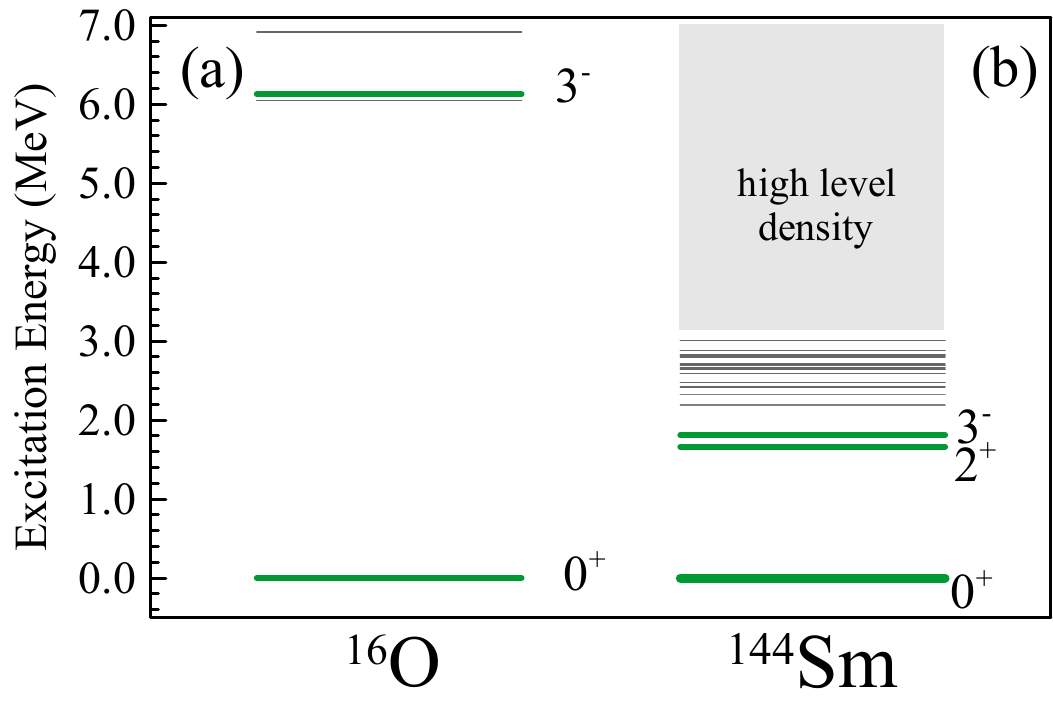}
\end{center}
\caption{The spectra of the $^{16}$O and $^{144}$Sm nuclei. Thick green lines represent 
energies of the intrinsic states considered in the CC calculation. The thin black line corresponds to neglected states.}
\label{spectra}
\end{figure}
First, we consider the $^{16}$O + $^{144}$Sm scattering at a collision energy just above the Coulomb
barrier ($V_{\rm B} = 61.4$ MeV). We evaluate the elastic angular distribution at $E_{\rm c.m.} = 62.3$ MeV
($0.9$ MeV above $V_{\rm B}$), by performing a CC calculation involving the lowest excited state of the 
$^{16}$O projectile, $\{J^\pi,\varepsilon^*\ {\rm (MeV)} \} = \{3^-,6.16\}$, and the main collective excitations 
of the target, $\{ 2^+,1.66\}$ and $\{3^-,1.81\}$, which have direct transitions to the ground state. 
Since the projectile is a doubly magic nucleus, 
its level density is very low, and the energy of the lowest excited state is very high. Then, the projectile's 
excitations are not expected to exert a strong influence on elastic scattering at near-barrier energies. On the other hand, the level density 
of the target is much higher. The spectra of the projectile and the target are shown in Fig.~\ref{spectra}(a) and
Fig.~\ref{spectra}(b), respectively. The levels taken 
into account and the levels neglected in the calculations are represented by thick green lines and thin gray lines, 
respectively. For the real and imaginary parts of the bare potential, $V$ and $W_{\rm F}$, we adopted the SPP 
and the short-range WS of Eqs.~(\ref{WF1}), (\ref{WF2}), and (\ref{WSpar}), respectively. Then, we evaluated the elastic cross section
and the polarization potential. \\

\begin{figure}
\begin{center}
\includegraphics[width=8.5 cm]{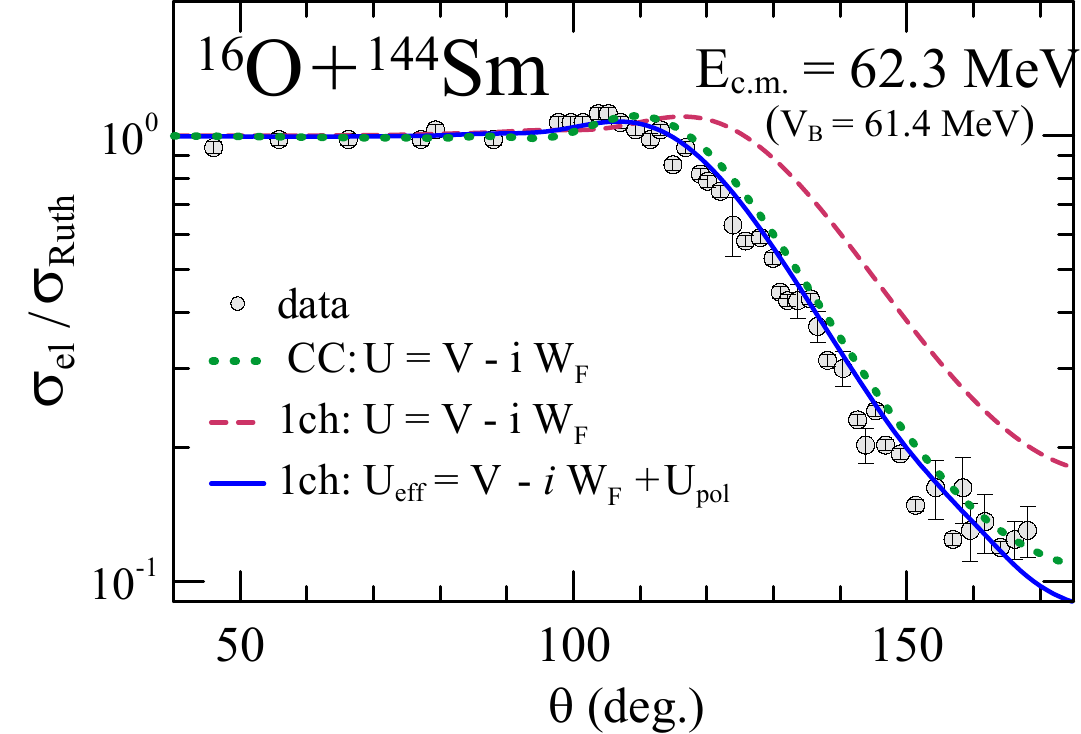}
\end{center}
\caption{The experimental elastic angular distribution of the $^{16}$O + $^{144}$Sm system at 
$E_{\rm c.m.} = 62.3$ MeV~\cite{ADT89}, normalized with respect to the Rutherford cross section. 
The data are compared to cross sections of different calculations. The real potential $V$ is the 
SPP for the $^{16}$O + $^{144}$Sm system, and $W_{\rm F}$ is the short-range WS function with 
the parameters of Eq.~(\ref{WSpar}).}
\label{sigel-over-sigruth}
\end{figure}
In Fig.~\ref{sigel-over-sigruth}, we compare the experimental elastic angular distribution of 
Abriola {\it et al.}~\cite{ADT89} with the results of our calculations. The dotted green line corresponds to the CC calculation, while the solid blue line and the dashed red line correspond to one-channel calculations with the potentials $U_{\rm eff}(R) = U(R) + U_{\rm pol}(R)$ and $U(R)$, respectively. First, we
observe that the dashed red line is quite different from the experimental cross section. The rainbow maximum
is shifted to the right, and the cross section at large angles is much higher than the data. This clearly shows
that channel couplings, neglected in this calculation, exert a strong influence on elastic scattering. However, 
the CC calculation describes the data extremely well. This means that the channel-coupling effects that influence the data can be attributed to the low-lying collective channels included in the calculation. Another interesting point is that the solid blue line and the dotted green lines are very close. This indicates that, in this case, the polarization potential of Thompson, Nagarajan, and Lilley~\cite{TNL89} can reproduce the 
channel-coupling effects on the elastic scattering data successfully.\\

\begin{figure}
\begin{center}
\includegraphics[width=8.5 cm]{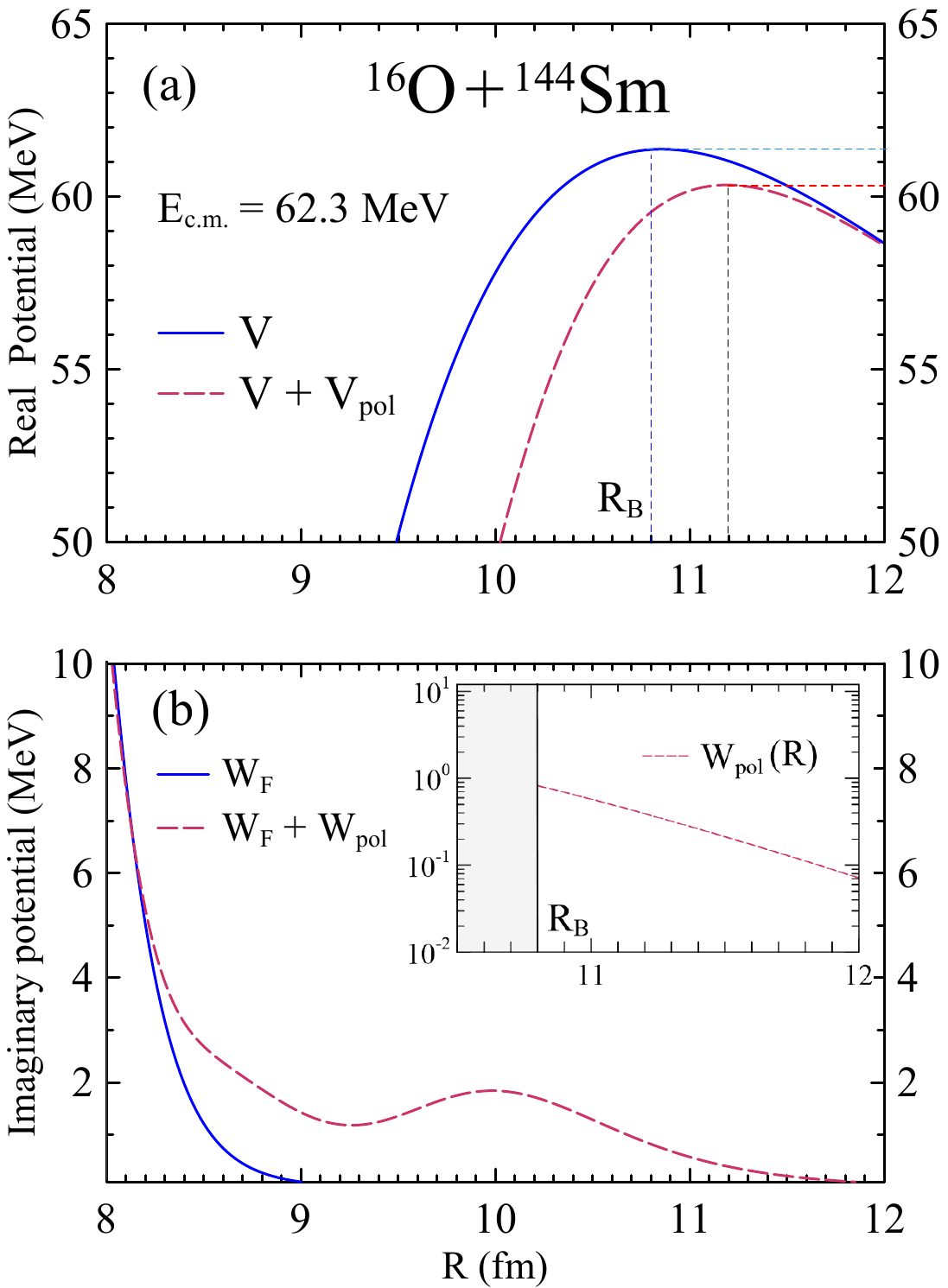}
\end{center}
\caption{The real (a) and the imaginary (b) parts of the potentials $U(R)$ and $U_{\rm eff}(R)$ used in the one-channel calculations of the previous figure, plotted in the neighborhood of the Coulomb 
barrier. Here, $V$ and $W_{\rm F}$ are the same potentials of the previous figure.}
\label{u-upol}
\end{figure}
The changes in the real and imaginary parts of the interaction arising from the contribution of the polarization potential are shown
in Fig.~\ref{u-upol}(a) and Fig.~\ref{u-upol}(b), respectively. The solid blue lines and the dashed red lines correspond to 
the components of $U(R)$ and $U_{\rm eff}(R)$, respectively. We find that the contribution of $V_{\rm pol}$
reduces $V_{\rm B}$ by $\sim 1$ MeV, increases $R_{\rm B}$ by $\sim 0.3$ fm, and makes the barrier thinner.
This leads to changes in phase shifts for $l=0$ and higher partial waves, which modify the interference 
pattern of the scattering amplitude. Consequently, appreciable changes are expected in the elastic cross 
section. \\

Comparing the imaginary parts of the two potentials, we find that the contribution of $W_{\rm pol}$ tends to be irrelevant at small distances, but increases the range of $W_{\rm eff}$.
Note that above $\sim 9$ fm, the absorption arises exclusively from $W_{\rm pol}$. It corresponds
to the part of the incident flux that is diverted to the inelastic channels included in the channel expansion.
This is expected to exert significant influence on the elastic cross section. It would also affect the fusion cross
section. A fraction of the incident flux would be absorbed by the polarization potential before reaching the strong-absorption interior region. Although this effect on the elastic wave function is correctly predicted by this 
one-channel 
calculation, it does not consider the contributions from the nonelastic channels to the fusion cross section. 
This is the reason the polarization potential approach tends to underestimate the fusion cross section. We will return to this issue in Sect. \ref{Fusion}. \\

We expect other inelastic channels to affect the reaction dynamics at higher collision energies. This could not be checked because the available data for  $^{16}$O + $^{144}$Sm fusion is restricted to energies very close to the 
Coulomb barrier. Instead, in the next section, we consider reactions for loosely bound projectiles. This offers an example of collisions where the breakup channels (excitations of the projectile) exert a 
strong influence on elastic scattering, but excitations of the target are also important. Since our calculations cannot
account for the two processes simultaneously, it is necessary to resort to polarization potentials.


\subsection{Elastic scattering of weakly bound systems}


Now we discuss the elastic scattering of a weakly bound projectile that breaks up into two clusters, denoted by  $c_1$ and  $c_2$. Owing to the low energy binding them, 
the breakup channel plays a major role in the reaction dynamics. Thus, the influence of the continuum must be taken into account. This can be done by the CDCC method, 
briefly discussed in Sect. \ref{Theory}. \\

We consider the elastic scattering of the $^8{\rm B}\, +\, ^{58}$Ni system, investigated in Ref.~\cite{LCA09}. This work analyzes the data of Aguilera {\it et al.}~\cite{AML09} at several 
bombarding energies. For the present discussion, we focus on the angular distribution for $E_{\rm lab} = 23.4$ MeV ($E_{\rm c.m.} = 20.6$ MeV), which is very close 
to the Coulomb barrier ($V_{\rm B} = 20.8$ MeV). It is basically the same energy region of the $^{16}$O$+^{144}$Sm collision investigated in the previous section. \\

The $^8$B projectile breaks up into a $^7$Be cluster and a proton, and the breakup threshold is only 137 keV.  The projectile-target interaction for the CDCC calculation
is written as
\begin{equation}
U({\bf R},{\bf r}) = U^{(1)}(r_1) \,+\, U^{(2)}(r_2),
\label{UPT}
\end{equation}
where ${\bf R}$ is the vector between the center of mass of the projectile and the target, $r_i$ is the distance
between the cluster $c_i$ and the target, and $\bf{r}$ is the vector joining the two clusters. These potentials have the form
\begin{equation}
U^{(i)}(r_i) = V^{(i)}(r_i)\,-i\,W^{(i)}(r_i),\ \ i=1,2,
\label{V-W(ri)}
\end{equation}
where the imaginary part of the potentials simulates the effects of the continuum and possibly of direct reaction channels left out of the model space. Note that this interaction is not diagonal in the channel space. We refer to the diagonal monople part as the channel potentials, and the off-diagonal or diagonal but monople part as the coupling interactions. \\

Taking into account excitations of the target in CDCC calculations is time-consuming and requires significant computational power. For this reason, typical CDCC calculations of elastic scattering, inelastic scattering, and breakup reactions consider the influence of neglected channels approximately. Instead of using the SPP plus short-range absorption, they adopt phenomenological optical potentials for the cluster-target 
interaction. They are selected by imposing the fact that they result in a good description of each cluster-target elastic scattering at the relevant incident energy. 
Thus, by construction, they must simulate the absorption resulting from transitions to neglected inelastic channels, which occur in grazing collisions. For this purpose,
the imaginary part of the phenomenological potentials must have a long range, reaching the barrier radius and beyond. \\

In the case of the $^8{\rm B}\, +\, ^{58}$Ni system, the two clusters are the proton ($c_1$) and the $^7$Be core ($c_2$).
For the $p-^{58}$Ni interaction, $U^{(1)}$, these authors adopted the complex potential proposed by Becchetti and Greenlees~\cite{EsB96}, which is widely used for collisions of protons with different nuclei. The $^7{\rm Be}$ - $^{58}{\rm Ni}$ interaction, $U^{(2)}$, was represented by the phenomenological complex potential of Moroz {\it et al.}~\cite{Mor82}. 
Then, the projectile target interaction, which we denote by $U_{\rm opt}$, is obtained by inserting these cluster-target potentials in Eq.~(\ref{UPT}). Finally, the p-$^{7}$Be states (bound and continuum) are generated using the Esbensen and Bertsch potential ~\cite{EsB96}. All details of these CDCC calculations can be found inRef.~\cite{LCA09}.\\

\begin{figure}
\begin{center}
\includegraphics[width=9 cm]{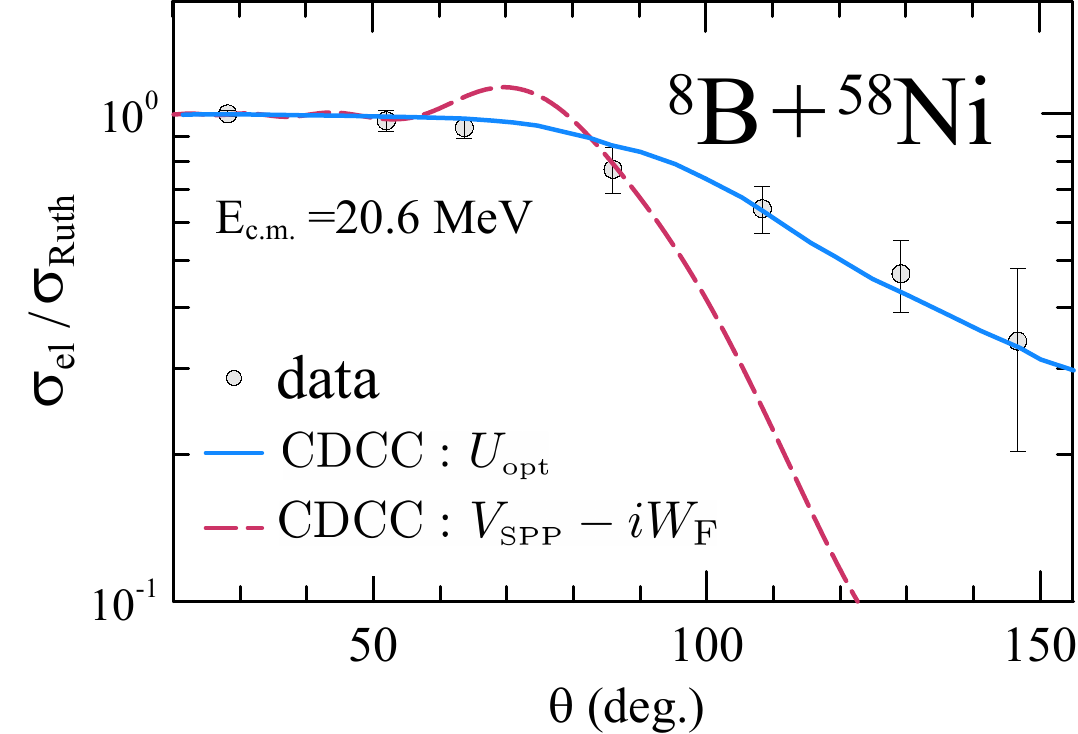}
\end{center}
\caption{The elastic scattering data of the $^8$B + $^{58}$Ni system~\cite{AML09} compared to predictions of two CDCC calculations. The solid line was obtained
with the phenomenological optical potential for the cluster-target interactions. Their sum is denoted by $U_{\rm opt}$. In the dashed red line, the real part of the cluster-target interactions is given by the SPP, and their imaginary parts are short-range WS functions. See the text for further details.}
\label{sigel_weakly bound}
\end{figure}
The model space of this calculation includes the elastic channel ($J^\pi = 3/2^-$) and a series of continuum discretized bins with energies up to 8 MeV. We consider bins with 
relative orbital angular momenta from 0 to $3\hbar$ between the clusters.\\

Figure~\ref{sigel_weakly bound} shows the elastic angular distribution at $E_{\rm c.m.} = 20.6$ MeV, measured by Aguilera {\it et al.}~\cite{AML09}, compared to the results of the above-mentioned CDCC calculation~\cite{LCA09} (solid blue line). Clearly, the theoretical cross section agrees very well with the data. This indicates that the phenomenological
cluster-target interactions simulate very well the effects of couplings to excitations of the target. \\

To assess the importance of couplings with intrinsic states of the target, we perform a CDCC calculation where they are neglected. Our calculations are very similar
to those of Ref.~\cite{LCA09}, but we use different interactions between the clusters and the target. For the real parts of $U^{(1)}$ and $U^{(2)}$, we adopt the SPP between the corresponding cluster and the target. For their imaginary parts, we use short-range WS functions, with the parameters of Eq.~(\ref{WSpar}). The projectile-target interaction
obtained by inserting these potentials into Eq.~(\ref{UPT}) is denoted by $V_{\rm SPP} -i W_{\rm F}$.\\

The angular distribution of the above CDCC calculation is represented by a dashed red line in Fig.~\ref{sigel_weakly bound}. The prediction does not agree with the data. The cross 
section at large angles is much lower, and it exhibits a pronounced rainbow maximum that is not observed in the data, nor predicted by the calculation with the 
phenomenological cluster-target potentials. This is due to the absence of the long range absorption in $W_{\rm F}$, which damps the cross section in grazing collisions, that scatter to the region of the rainbow maximum.


\section{Fusion reactions}\label{Fusion}



If the polarization potential is realistic, it is expected that the single-channel elastic cross section using such a potential
will be very close to the data and to the one obtained in the full CC calculation. \\
However, the situation is entirely different when one considers the fusion cross section. In a typical calculation with a strong imaginary potential with a short range, $W_{\rm F}(R)$, the fusion cross section takes the form~\cite{CaH13},
\begin{equation}
\sigma_{\rm F} = \sum_{\alpha = 0}^{n-1}\ \sigma_{\rm F}^{(\alpha)}
\label{sigF-CC1}
\end{equation}
with
\begin{equation}
\sigma_{\rm F}^{(\alpha)} = \frac{k}{E}\ \left< \psi_\alpha\left|\ W_{\rm F}\ \right| \psi_\alpha \right> .
\label{sigF-CC2}
\end{equation}
Carrying out angular momentum projection, we get
\begin{equation}
\sigma_{\rm F}^{(\alpha)} = \frac{\pi}{k^2}\ \sum_l (2l+1)\ P_{\rm F}^{(\alpha)}(l),
\label{sigF-alpha}
\end{equation}
where $P_{\rm F}^{(\alpha)}(l)$ is the fusion probability in channel$-\alpha$ in a collision with angular momentum $l$. It is given by,
\begin{equation}
P_{\rm F}^{(\alpha)}(l) = \frac{4k}{E}\ \int dR\ \left| u_{\alpha,l} (R) \right|^2 \ W_{\rm F}(R).
\label{PF-alpha-l}
\end{equation}
where $u_{\alpha,l} (R)$ is the radial wave function in channel $\alpha$ at the $l^{th}$ partial-wave. For simplicity, we neglected spins in the angular-momentum expansion.
The inclusion of spin introduces just a few geometrical factors.\\

In an ideal CC problem, the channel expansion includes all direct channels that affect the reaction dynamics. Then, the 
diagonal nucleus-nucleus interaction is the sum of a bare real potential, like the SPP, with $W_{\rm F}$. However, in typical heavy-ion collisions, the spectra of the collision partners may have a large number 
of states, each with complicated nuclear structure. This leads to large uncertainties in their coupling matrix elements, which preclude them from being included in the CC model space. \\

\subsection{Fusion of tightly bound systems}

We perform CC calculations for the $^{16}$O + $^{144}$Sm system using the same model space of the elastic scattering calculation of the previous section.
That is, we include the elastic channel ($|0)$) and the main inelastic channels influencing the collision dynamics, namely the quadrupole vibration with $J^\pi = 2^+, \varepsilon^* = 1.66$ MeV  ($|1)$), and the octupole vibration with $J^\pi = 3^-, \varepsilon^* = 1.81$ MeV ($|2)$) of the target, and the $J^\pi = 3^-, \varepsilon^* = 6.13$ MeV 
of the projectile ($|3)$).  The $P$ and the $Q$ projectors are then given by,\\
\begin{equation}
P = \left| 0 \right)\,\left( 0 \right|; \ \ \  Q = \left| 1 \right)\,\left( 1 \right|\,+\, \left| 2 \right)\,\left( 2 \right| \,+\, \left| 3 \right)\,\left( 3 \right|.
\end{equation}
For the real and imaginary parts of the projectile-target interaction, we adopted the SPP~\cite{CPH97,CCG02} and a WS function with the 
short-range parameters of Eq.~(\ref{WSpar}), respectively. \\

\begin{figure}
\begin{center}
\includegraphics[width=8.5 cm]{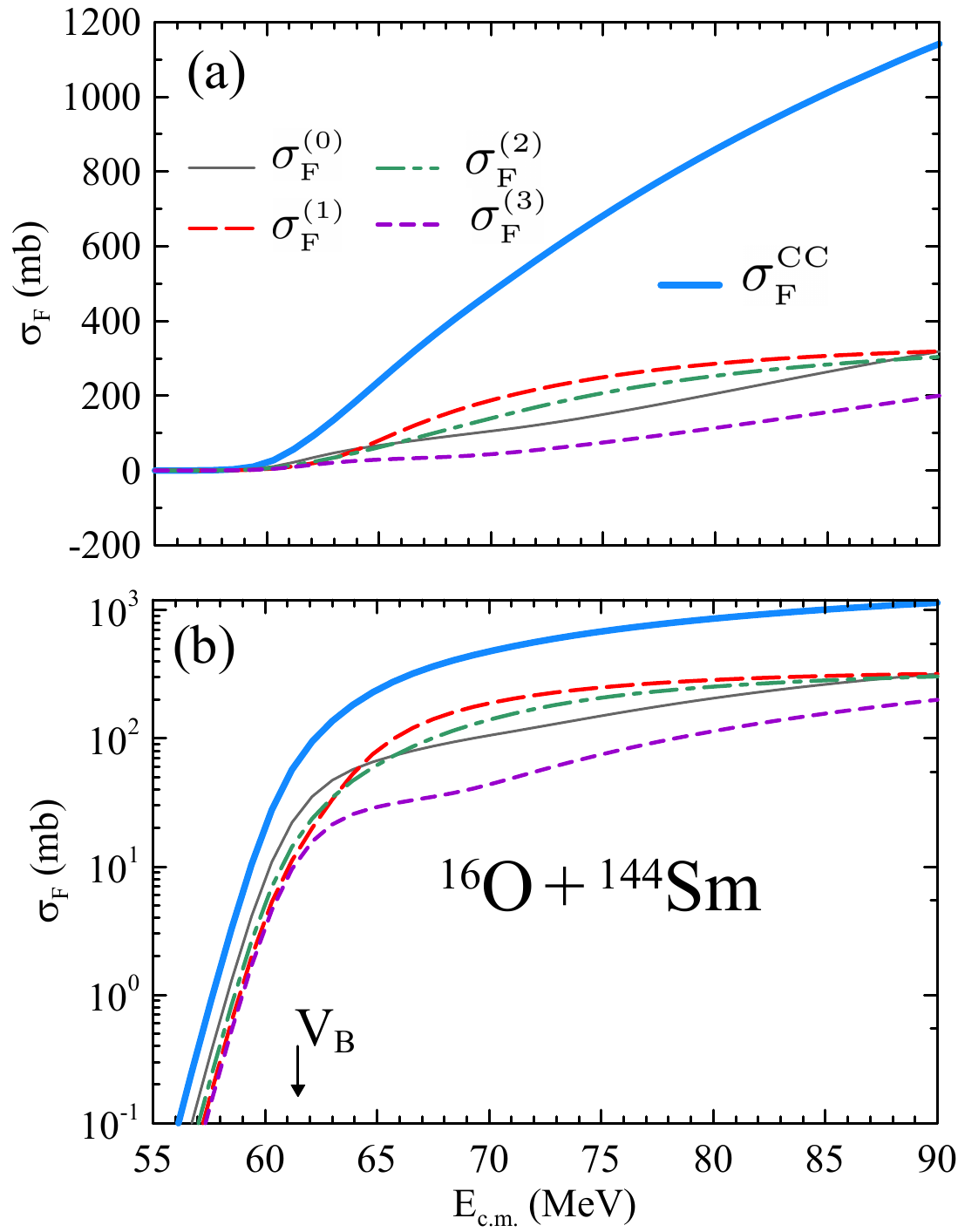}
\end{center}
\caption{The CC fusion cross section of the $^{16}$O + $^{144}$Sm system
($\sigma_{\rm F}^{\scriptscriptstyle\rm CC})$, and the contribution from the elastic 
($\sigma_{\rm F}^{\scriptscriptstyle (0)})$, and from the three inelastic
channels in the model space ($\sigma_{\rm F}^{\scriptscriptstyle (1)}$, $\sigma_{\rm F}^{\scriptscriptstyle (2)}$, and $\sigma_{\rm F}^{\scriptscriptstyle (3)}$), given by Eqs.~(\ref{sigF-CC1}), (\ref{sigF-CC2}), 
and (\ref{sigF-alpha}).} 
\label{sigF-components}
\end{figure}
The CC fusion cross section is shown in Fig.~\ref{sigF-components}, together with the contributions from the 
elastic and the three inelastic channels included in the calculation.  As we have shown at the beginning of this section, $\sigma_{\rm F}^{\rm CC}$ can be written as the sum of contributions from the channels included
in the model space, namely 
$\sigma_{\rm F}^{\scriptscriptstyle(0)}, \sigma_{\rm F}^{\scriptscriptstyle(1)}, \sigma_{\rm F}^{\scriptscriptstyle(2)}$,
and $\sigma_{\rm F}^{\scriptscriptstyle(3)}$ (see Eqs.~(\ref{sigF-CC1})
and (\ref{sigF-CC2})). Comparing $\sigma_{\rm F}^{\scriptscriptstyle\rm CC}$ with $\sigma_{\rm F}^{\scriptscriptstyle (0)}$, one finds that the contribution of the elastic channel is only $\sim 25-30\%$ of $\sigma_{\rm F}^{\rm CC}$. The remaining part comes from the fraction of the incident current that was diverted to the inelastic channels.\\

Now we consider the fusion cross section in the polarization potential approach. First, we determine the elastic wave function, $\psi$, by solving Eq.~(\ref{eff-1ch}), with the effective potential of Eqs.~(\ref{Ueff}),
(\ref{U}), and (\ref{Upol}). In calculations with a short-range imaginary potential, the absorption cross section corresponds to fusion. Here, the situation is more complicated. We can consider the following absorption cross sections, \\
\begin{eqnarray}
\sigma_{\rm abs}^{\rm min} &=& \frac{k}{E}\,\left< \psi \left|\,W_{\rm F}\,
\right|\psi \right>, \label{sigmin0}\\
\sigma_{\rm abs}^{\rm max} &=& \frac{k}{E}\,\left< \psi \left|\,W_{\rm F}
+W_{\rm pol}\,\right|\psi \right>. \label{sigmax0}
\end{eqnarray}

The relation between the above cross sections and $\sigma_{\rm F}^{\rm CC}$ is not clear. To shed light on this issue, we consider the evolution of the 
incident elastic current in a qualitative picture. 
As the projectile approaches the target, the incident current is attenuated
by the action of the coupling interaction, which diverts a fraction of it
to the inelastic channels. When the elastic current reaches the barrier, it is partly transmitted, and the transmission determines the contribution 
of the elastic channel to $\sigma_{\rm F}^{\rm CC}$. This contribution is given
by Eq.~(\ref{sigF-CC2}), for $\alpha=0$. Keeping in mind that the
inclusion of the polarization potential in the one-channel equation 
should lead to an excellent approximation to the elastic channel of the CC calculation, Eqs.~(\ref{sigmin0}) and (\ref{sigF-CC2}) for $\alpha = 0$ should be equivalent. Then, we can write \\
\begin{equation}
\sigma_{\rm abs}^{\rm min} \simeq \sigma_{\rm F}^{(0)}.   
\label{sigmin}
\end{equation}

One must also follow the evolution of the currents in the inelastic channels. 
When they reach the barrier, they are partially transmitted and partially 
reflected. The transmitted part contributes to the fusion cross section, while
the reflected part gives the inelastic cross section of the corresponding 
channel. For a proper calculation of the fusion and inelastic cross sections,
one needs the transmission rate for each channel. These rates are determined 
in the CC calculations since they provide the wave functions of all channels
in the model space. However, these rates are no longer available in
the polarization potential approach, which only gives the elastic wave function. In this way, one can
only get a cross section for the sum of the inelastic and fusion processes. This
corresponds to the total reaction cross section, $\sigma_{\rm R}$. Then, 
we can write,\\
\begin{equation}
\sigma_{\rm abs}^{\rm max} \simeq \sigma_{\rm R}.   
\label{sigmax}
\end{equation}

\begin{figure}
\begin{center}
\includegraphics[width=8.5 cm]{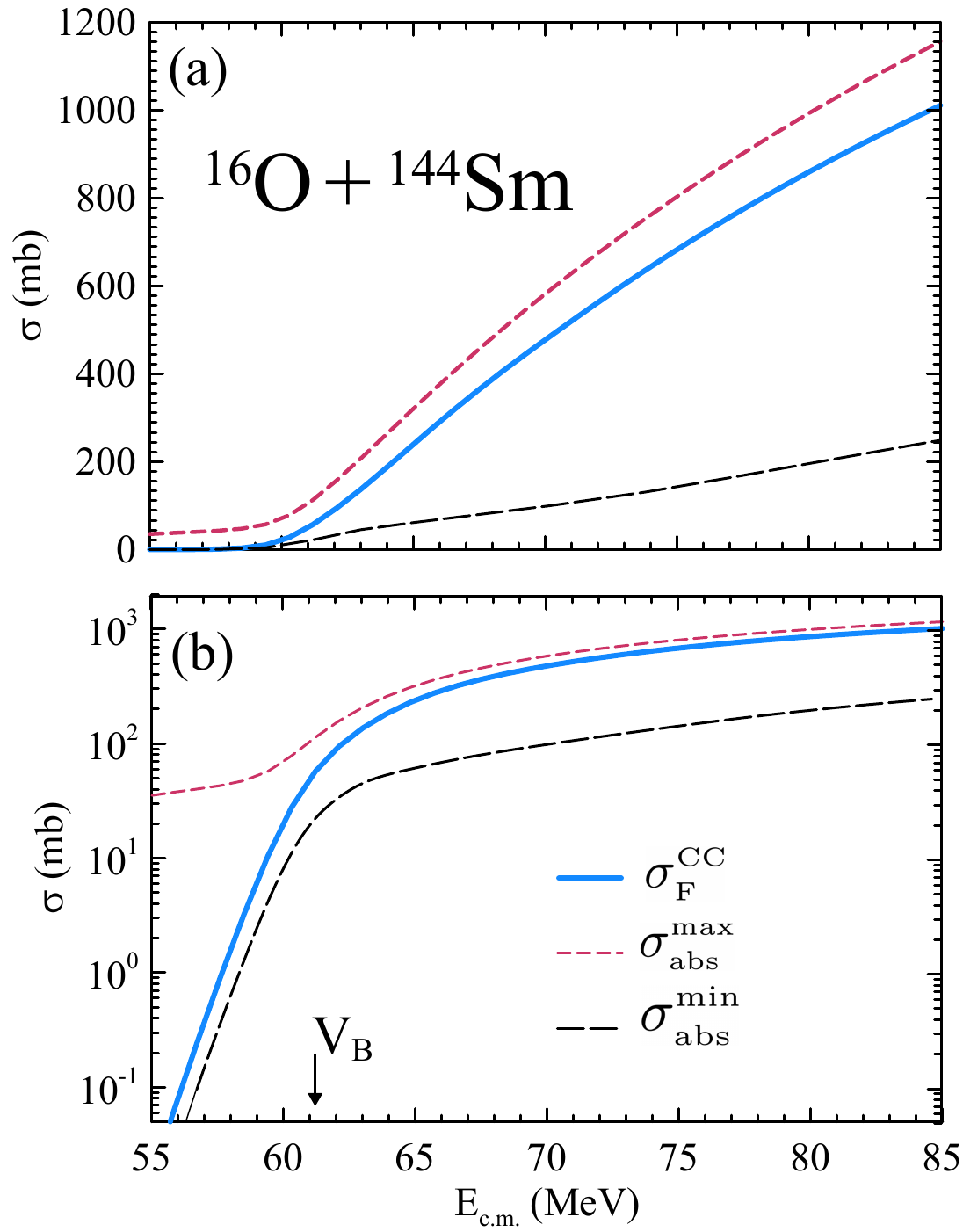}
\end{center}
\caption{The fusion cross section for $^{16}$O+$^{144}$Sm (panel (a) linear plot and panel (b) log plot): results for the CC calculation (thick blue line) and the limiting values of the absorption cross sections in the polarization potential approach, 
$\sigma_{\rm abs}^{\rm min}$ (black long-dashed line) and 
$\sigma_{\rm abs}^{\rm max}$ (red short-dashed line). See the text for further details.}
\label{sigmin-sigmax}
\end{figure}
Fig.~\ref{sigmin-sigmax} shows the two limiting values of the absorption cross
section in the polarization potential approach, together with the fusion cross
section of the CC calculation. As we have shown, $\sigma_{\rm abs}^{\rm min}$ 
is roughly equal to the contribution of the elastic channel to the fusion cross 
section. Since  $\sigma_{\rm F}^{\rm CC}$
contains important contributions from inelastic channels (see Fig.~\ref{sigF-components}),
$\sigma_{\rm abs}^{\rm min}$ is much smaller than it. \\

On the other hand, $\sigma_{\rm abs}^{\rm max}$ is much larger than the fusion
cross section of the CC calculation. Above the Coulomb barrier, the difference
is $\sim 20\,\%$. However, at sub-barrier energies, it is orders of magnitude
larger. This is due to the long range of the Coulomb quadrupole coupling.
Although its strength decreases as the classical turning point
increases, the tunneling probability decreases much faster. For this reason, 
the ratio $\sigma_{\rm R}/\sigma_{\rm F}^{\rm CC}$ increases rapidly at 
sub-barrier energies.\\

\begin{figure}
\begin{center}
\includegraphics[width=8.5cm]{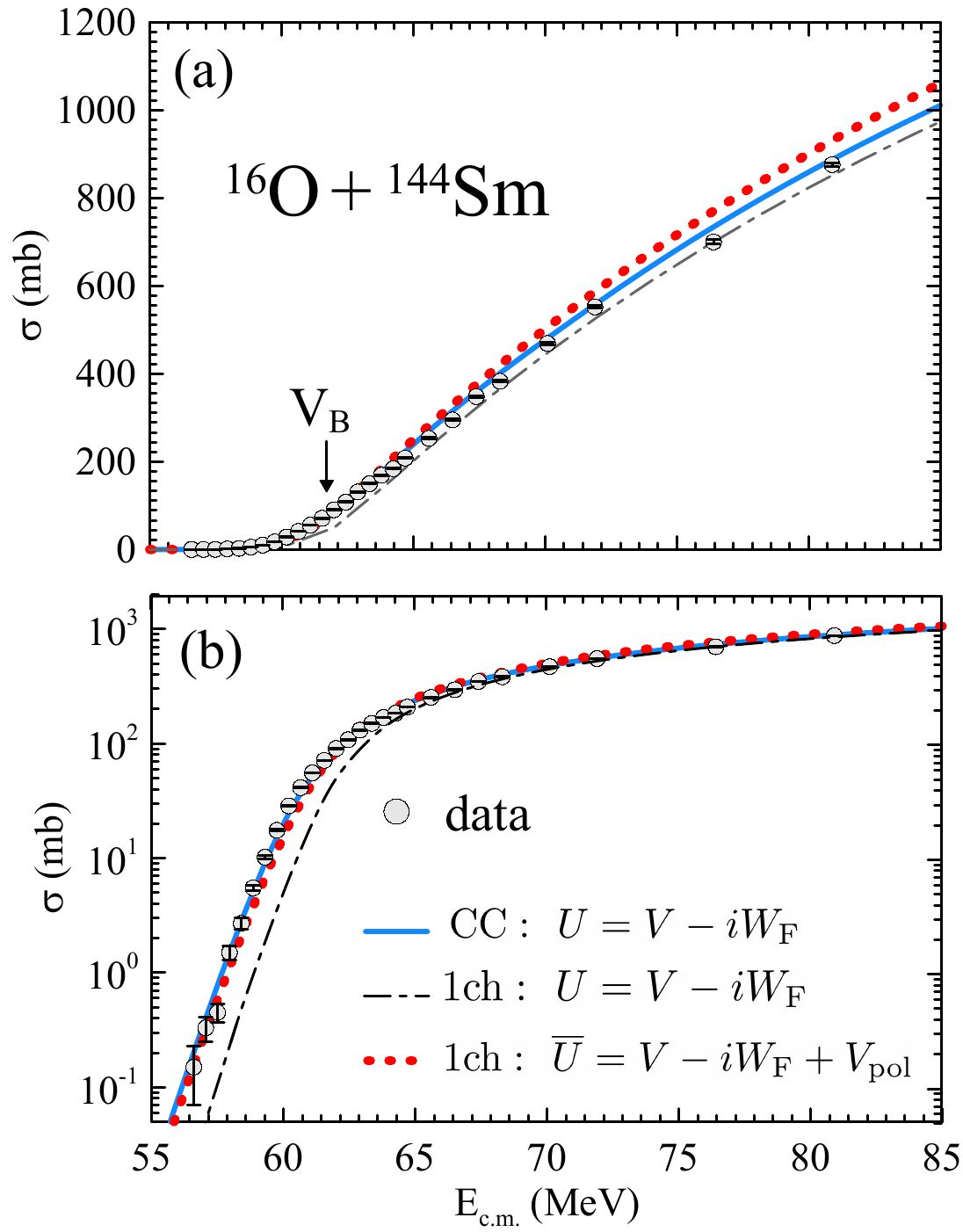}
\end{center}
\caption{The experimental fusion data of Abriola {\it et al.}~\cite{ADT89} and results of different 
calculations. The kind of calculation (CC or 1 channel) and the adopted potentials are indicated in 
the legend. As explained in the text, $V$ is the SPP for the system, $W_{\rm F}$ is the short-range 
imaginary potential of Eqs.~(\ref{WF1}), (\ref{WF2}) and (\ref{WSpar}), and $V_{\rm pol}$ is the real
part of the polarization potential. Panel (a)  is the linear plot, and panel (b) is the log plot.}
\label{app-vs-data_1}
\end{figure}
Figure~\ref{app-vs-data_1} compares the experimental cross section of Abriola {\it et al.}~\cite{ADT89} with the results of different calculations. The potential adopted in each calculation is indicated in the legend 
and the caption. Clearly, $\sigma_{\rm F}^{\rm CC}$ (solid blue lines) is in excellent agreement with the data 
at all collision energies. Further, we find that the cross section of the one-channel calculation with $U(R)$,
where all couplings are neglected (thin black dot-dashed line), is close to the data at above-barrier energies but much lower at sub-barrier energies. This is a well-known effect. Channel couplings lead to a lower effective 
barrier~\cite{DLW83,DLW83a}. \\

The barrier-lowering effect can be simulated by adding the polarization 
potential to the interaction. That is, 
\[
U(R)\ \rightarrow \ U_{\rm eff}(R) = U(R) + U_{\rm pol}(R),
\]
with
\begin{equation}
U_{\rm pol}(R) = V_{\rm pol}(R)-iW_{\rm pol}(R).
\label{Ubar}
\end{equation}
As shown in Fig.~\ref{u-upol}(a), $V_{\rm pol}(R)$ is negative around the barrier radius. Then, the addition of
$V_{\rm pol}$ to the real part of the interaction lowers the barrier by 1 MeV.
However, $U_{\rm pol}$ also contains an imaginary part, $W_{\rm pol}$, which absorbs the incident wave. Then, 
two limiting absorption cross sections can be evaluated, $\sigma_{\rm abs}^{\rm min}$ and 
$\sigma_{\rm abs}^{\rm max}$ (see Eqs.~(\ref{sigmin0}) and (\ref{sigmax0})). As shown in Fig.~\ref{sigmin-sigmax}, 
neither is close to $\sigma_{\rm F}^{\rm CC}$. The former is much lower, while the latter is much higher. \\

Then, we tried a different approximation. We considered $V_{\rm pol}$ but dropped $W_{\rm pol}$ 
(dotted red line). That is, 
\begin{equation}
U(R)\ \rightarrow \ \overline{U}(R) = \overline{V}(R) - i\, W_{\rm F}(R),
\label{Ubar}
\end{equation}
with
\begin{equation}
\overline{V}(R) = V(R)+V_{\rm pol}(R).
\label{Vbar}
\end{equation}
We performed a one-channel calculation with the above potential, and the resulting fusion cross section is
represented by the dotted red lines in Fig.~\ref{app-vs-data_1}. The agreement with the full  CC calculation and with the data is excellent. This curve can hardly be distinguished from the solid blue line.\\

\begin{figure}
\begin{center}
\includegraphics[width=8cm]{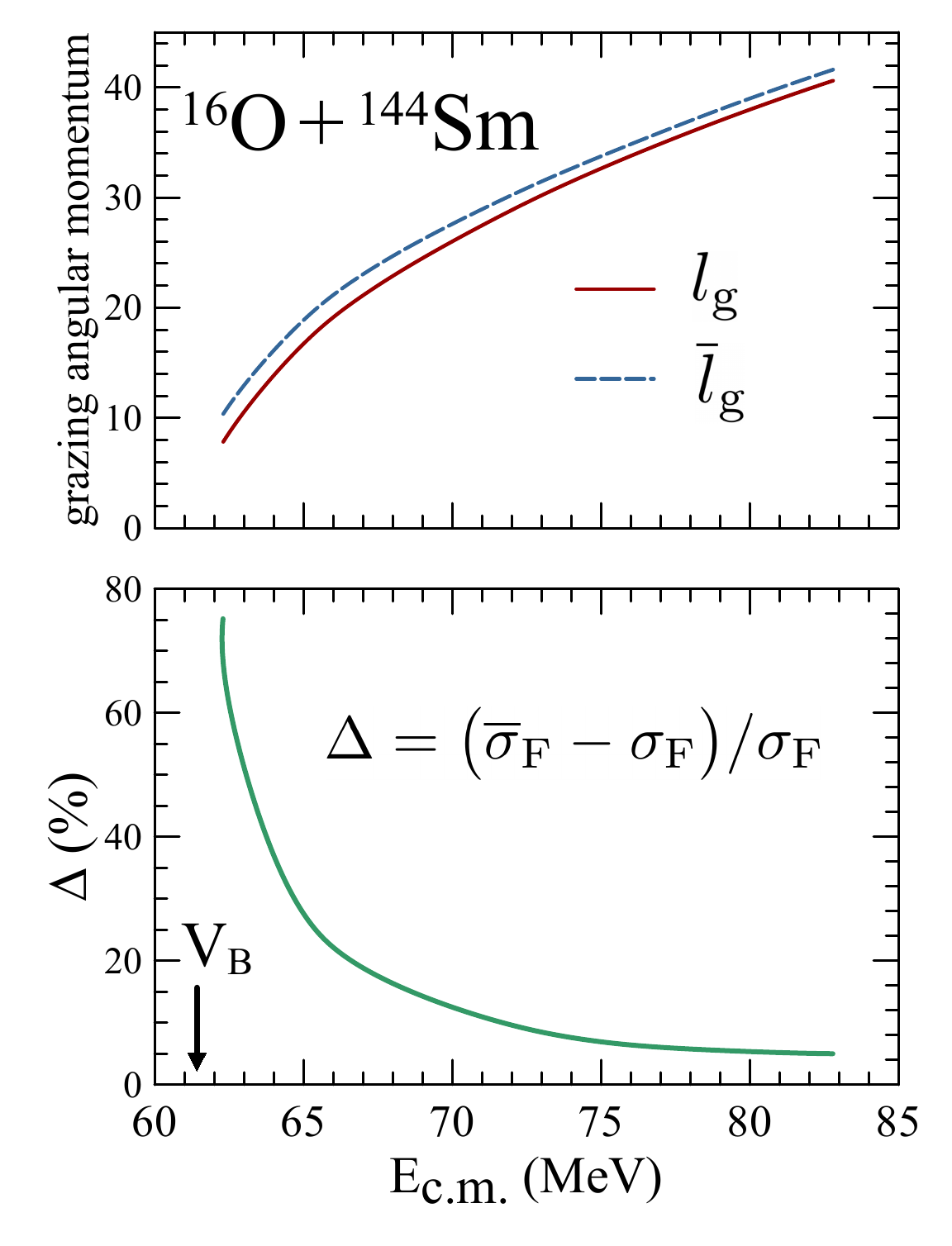}
\end{center}
\caption{The influence of $V_{\rm pol}$ on the grazing angular momentum and the fusion cross section. 
Using $V_l(R)$ and $V_l(R)+V_{\rm pol}(R)$, we obtained respectively the grazing angular
momenta $l_{\rm g}$ and $\overline{l}_{\rm g}$, and the corresponding fusion cross sections $\sigma_{\rm F}$
and $\overline{\sigma}_{\rm F}$.}
\label{vpol-sigF}
\end{figure}
It is very interesting that the results of the one-channel calculation that neglects 
all channel-coupling effects are close to $\sigma_{\rm F}^{\rm CC}$ at above-barrier 
energies. The reason is that the polarization potential in this energy region is very 
weak. Then, the resulting grazing angular momentum is only slightly modified. Consequently,
the fusion cross section, which is accurately given by Eq.~(\ref{sigF-lg}), is not significantly
changed. \\

The influence of the polarization potential is illustrated in Fig.~\ref{vpol-sigF}. The grazing angular momenta associated with 
$\overline{V}(R)$ and $V(R)$ for the $^{16}$O + $^{144}$Sm system are shown in Fig.~\ref{vpol-sigF}(a). They are denoted 
by $\overline{l}_{\rm g}$ and $l_{\rm g}$, respectively. The two angular momenta are very 
close, and the difference between them decreases as $E_{\rm c.m.}$ increases. 
Using Eq.~(\ref{sigF-lg}), we can evaluate the influence of this difference on the fusion cross section. We consider
the relative change,
\begin{equation}
\Delta = \frac{\overline{\sigma}_{\rm F} - \sigma_{\rm F}}{\sigma_{\rm F}} = \frac{\overline{l}_g^{\,2} - l_g^2}{l_g^2},
\label{Delta}
\end{equation}
where $\overline{\sigma}_{\rm F}$ and $\sigma_{\rm F}$ are the fusion cross sections considering and not considering $V_{\rm pol}$.
Figure~\ref{vpol-sigF}(b) shows the correction $\Delta$ as a function of the collision energy. Although it nearly doubles 
the cross section just above the Coulomb barrier, its importance rapidly decreases as $E_{\rm c.m.}$ increases. At the 
highest energies in the plot, it is less than $5\%$.\\


\subsection{Fusion of weakly bound systems}


Short-range imaginary potentials are adopted in most calculations of fusion cross sections of weakly bound 
systems~\cite{DiT02,DTB03,MCD14,KCD18,RCL20,CRF20,LFR22,FRL23,CZF23}. In this section, we discuss this issue in 
the context of a recently developed CDCC-based method to evaluate CF and ICF cross 
sections~\cite{RCL20,CRF20,LFR22} in collisions of projectiles that dissociate into two clusters, $c_1$ and $c_2$ 
(see Refs.~\cite{CRF20,LFR22} for details).\\

In this method, the nucleus-nucleus interaction is given by the sum of two complex cluster-target potentials, 
$U^{(1)}$ and $U^{(2)}$. The real part of these potentials is the SPP, while their imaginary parts are given by short-range WS functions, with the parameters of Eq.~(\ref{WSpar}).\\

The method is implemented in two steps. In the first, the CDCC wave functions in the bound (${\rm \Psi_{B}}$) and the continuum-discretized (${\rm \Psi_{C}}$) sub-spaces (see Eq. (\ref{Psi-B_Psi-C})are evaluated by the FRESCO code~\cite{Tho88}.
Next, the angular momentum projected components of ${\rm \Psi_{B}}$ are used to evaluate expectation values
of the total imaginary potential. In this way, one gets the probability of direct fusion of the whole projectile (DCF) with the target at each
angular momentum. Similarly, the angular momentum projected components of ${\rm \Psi_{C}}$ are used to evaluate expectation values of $W_{\rm F}^{(1)}$ and $W_{\rm F}^{(2)}$, which give the inclusive probabilities of fusion of the clusters $c_1$ and $c_2$ with the target, respectively. They are the probabilities of fusion of one of the clusters, independently of what happens to the other.\\

However, the above probabilities do not correspond to the cross sections measured in actual experiments. They
can only measure the CF cross section, which corresponds to the sum of the DCF with the sequential fusion of the two clusters (SCF), following the breakup process, and the ICF cross sections. The latter corresponds to the process where only one of the clusters fuses with the target. Then, to relate the available theoretical cross section with the measured ones, some complementary assumptions are needed. This is the second step of the method. The probabilities for the calculated and the measured fusion processes are related by classical probability theory, as described in Ref.~\cite{LFR22}.\\

\begin{figure}
\begin{center}
\includegraphics[width=9cm]{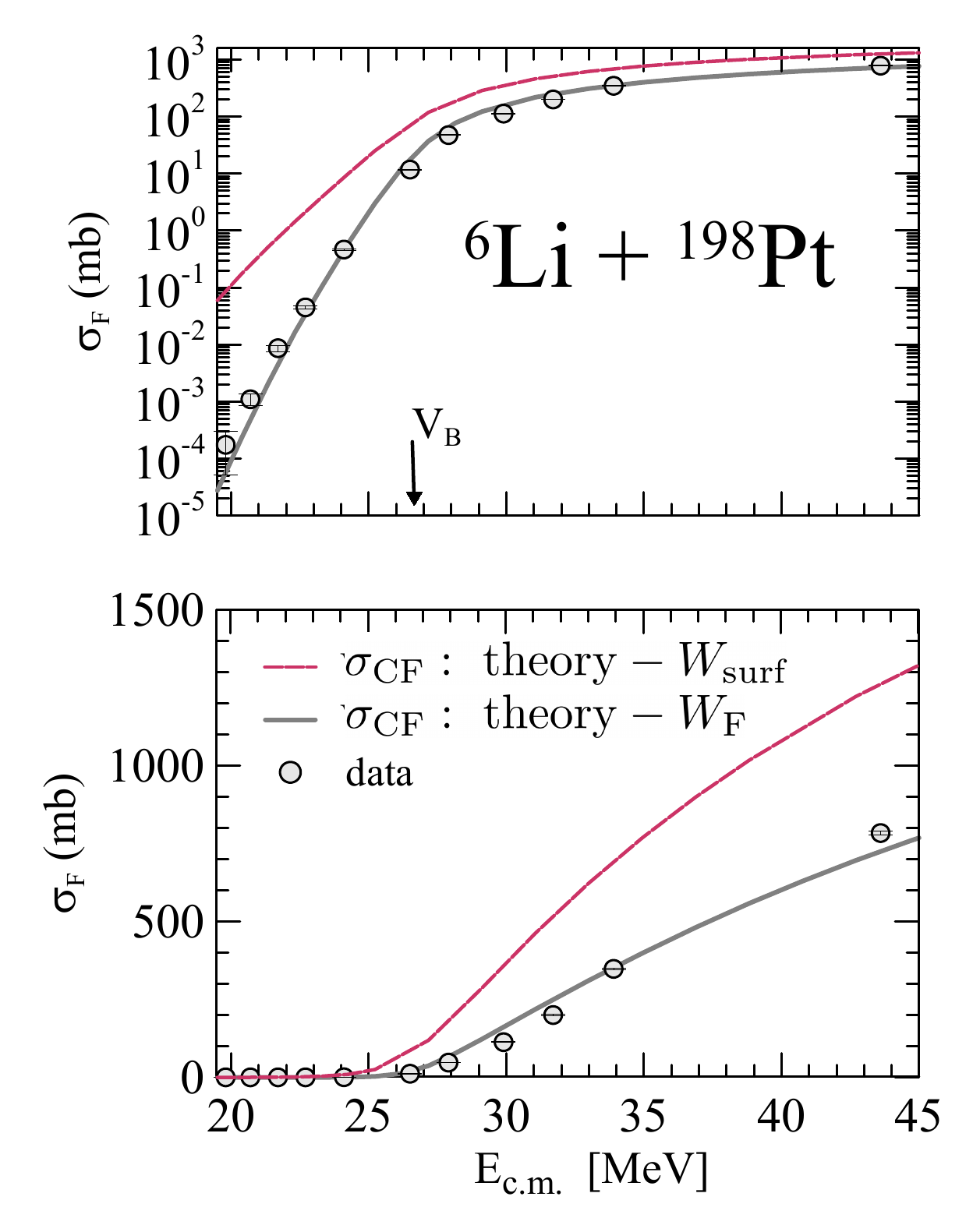}
\end{center}
\caption{The theoretical cross sections of Ref.~\cite{LFR22} compared to the data of 
Shrivastava {\it et al.}~\cite{SNL09}. The gray solid line and the dashed red line are the results of calculations
with the SPP, with short-range absorption ($W_{\rm F}$) and an imaginary potential reaching the surface 
($W_{\rm surf}$), respectively.}
\label{6Li-198Pt}
\end{figure}
The method was successfully applied to collisions of $^{6,7}$Li with $^{124}$Sn, $^{197}$Au, $^{198}$Pt, $^{209}$Bi, and $^{90}$Zr 
targets~\cite{RCL20,CRF20,LFR22, ZJZ24,CZF23},  and to collisions of $^6$He and $^{11}$Be with $^{208}$Pb and $^{238}$U~\cite{FRL23}. In the case of  $^6$He, the calculations were performed within the di-neutron approximation. 
The good agreement between predictions of this method and the data is illustrated in Fig.~\ref{6Li-198Pt},
which shows CF cross sections for the $^{6}$Li + $^{198}$Pt system. \\

Figure~\ref{6Li-198Pt} shows the theoretical CF cross section of Ref.~\cite{LFR22}, compared to the data 
of Shrivastava {\it et al.}~\cite{SNL09}. The solid gray line was obtained with the SPP and the short-range 
imaginary 
potential of Eqs.~(\ref{WF1}) and (\ref{WF2}), with the parameters of Eq.~(\ref{WSpar}). The model space 
and other details of the calculation can be found in Ref.~\cite{LFR22}. However, it is worth mentioning 
that the convergence of the partial-wave series of the fusion cross section is much faster than in typical
calculations of elastic and breakup cross sections. We got excellent convergence for 
$J_{\rm max} \sim 30$. Although much larger angular momenta affect elastic scattering and direct reactions,
the transmission coefficients through the corresponding potential barriers are vanishingly small. \\

Comparing theory and experiment, we conclude that the model of Ref.~\cite{LFR22} with the SPP and short-range absorption describes the CF data extremely well, above and below the Coulomb barrier. Note that the data 
extend to very low energies, reaching almost 7 MeV below $V_{\rm B}$. The good agreement in this region calls for further discussion. \\

In the comparison of $^{16}$O + $^{144}$Sm fusion data with CC cross sections at 
above-barrier energies (see Fig.~\ref{app-vs-data_1}), the agreement was also good.  However, the 
experimental cross section below the Coulomb barrier was strongly enhanced with respect to the theoretical 
prediction. This results from a barrier-lowering effect arising from inelastic couplings. Here, couplings 
with intrinsic states of the projectile are explicitly taken into account in the CDCC calculation.
However, couplings with excitations of the target are neglected. Then, the good agreement in this region
suggests that such couplings are irrelevant. To check this point, we performed a CC calculation ignoring 
the cluster structure of the projectile, but including the main collective states of the $^{198}$Pt target. 
The results were compared with those of a one-channel calculation neglecting all couplings. The two cross sections were essentially the same. They differed by less than $1\%$ at all collision 
energies. Owing to the low projectile charge, the Coulomb excitation that dominated the reaction dynamics 
of the $^{16}$O + $^{144}$Sm system at sub-barrier energies is very weak in collisions of $^6$Li projectiles.\\

Now we check the use of long-range absorption in the CF calculations. We perform a similar calculation by replacing $W_{\rm F}$ with an imaginary potential with a more extended range, $W_{\rm surf}$, reaching the surface. 
For simplicity, we kept the WS form but adopted larger radius and diffusivity parameters, namely
\[
W_0 = 37.0\ {\rm MeV},\ r_{\rm w} =1.18\ {\rm fm}, a_{\rm w}= 0.62\ {\rm fm}.
\]
They correspond to the parameters of the Aky\"uz-Winther potential (AW)~\cite{BrW04,AkW81}, with an
attenuation factor of 0.78. Adopting real and imaginary potentials with the same functional form is a 
simple way to study heavy-ion collisions. Alvarez {\it et al.}~\cite{ACH03} obtained reasonable 
descriptions of elastic scattering data using this procedure for several collision energies and
systems in a broad mass range. 
The CF cross section calculated with the SPP and this imaginary potential is represented by dashed red 
lines in Fig.~\ref{6Li-198Pt}. In this case, the theoretical cross section is systematically higher than the data.
At the lowest sub-barrier energies, it exceeds the CF data by several orders of magnitude. The high 
values in this energy region can be traced back to the tail of $W_{\rm surf}$, which simulates the influence 
of direct reaction channels left out of the model space. This is shown more clearly in the next figure.\\

\begin{figure}
\begin{center}
\includegraphics[width=8cm]{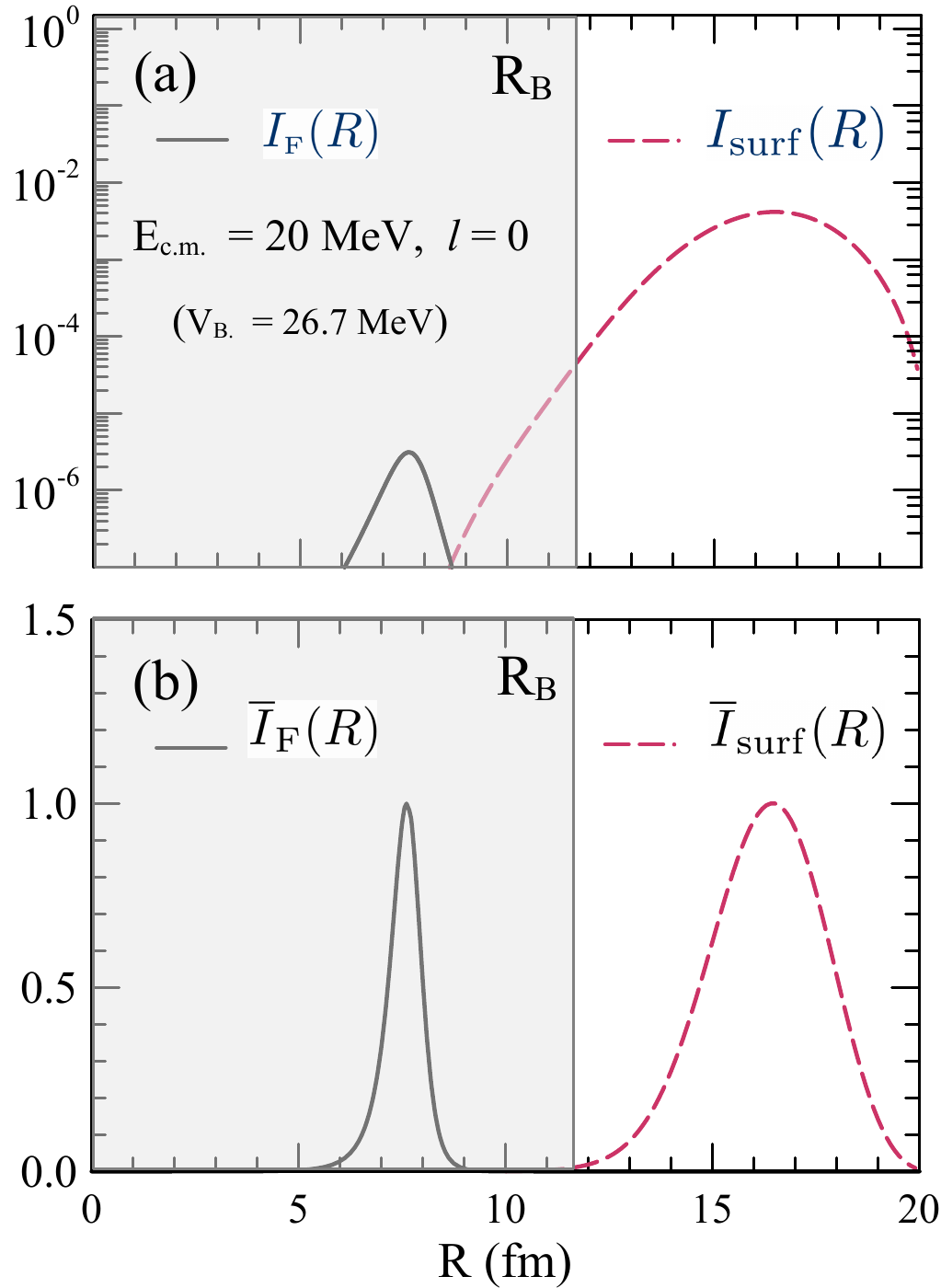}
\end{center}
\caption{The integrand of Eq.~(\ref{Irad}) for the imaginary potentials $W_{\rm F}(R)$ 
and $W_{\rm surf}(R)$. In (a) they are shown in a logarithmic scale. In (b), they are
normalized with respect to their maxima.}
\label{WF-Wsurf}
\end{figure}
The fusion cross section in a CC problem is calculated by Eqs.~(\ref{sigF-CC1}) to 
(\ref{PF-alpha-l}). At very low energies, the main contribution comes from low 
partial waves in the elastic channel. We then look at the probability
\begin{equation}
P_{\rm F}^{(0)}(l=0) = \frac{4k}{E}\ \int  I(R)\ dR,
\label{P00}
\end{equation}
where $I(R)$ is the integrand 
\begin{equation}
I(R) = \left| u_{0,0} (R) \right|^2 \ W(R),
\label{Irad}
\end{equation}
which establishes  the dominant $R$-region in the CF calculation. \\

Figure~\ref{WF-Wsurf} shows the integrands $I_{\rm F}(R)$ and $I_{\rm surf}(R)$ for the calculations with the imaginary 
potentials $W_{\rm F}(R)$ and $W_{\rm surf}(R)$. They were evaluated for a very low collision energy, 
$E = 20$ MeV, at the main partial-wave $l=0$. Comparing the two curves in  Fig.~\ref{WF-Wsurf} (a), 
one clearly sees that the maximum of $I_{\rm surf}(R)$ is several orders of magnitude higher than that 
of $I_{\rm F}(R)$. For this reason, 
the cross section obtained with the former is much higher than that obtained with the 
latter. Further, the maxima of the two functions are located at different $R-$regions. 
This is seen more clearly in Fig.~\ref{WF-Wsurf} (b), where these functions 
are normalized with respect to their maxima ($\overline{I}_{\rm F}(R)$ and 
$\overline{I}_{\rm surf}(R)$). In the case of $I_{\rm F}(R)$, absorption takes place
exclusively in the inner region of the Coulomb barrier. Thus, it simulates the fusion
process. On the other hand, $I_{\rm surf}(R)$ is only relevant for 
$R>R_{\rm B}$. Thus, the absorption simulates the attenuation of the ingoing component
of the elastic current that is diverted to the direct reaction channels neglected
in the model space. Therefore, the large cross section at low energies corresponds
to inelastic scattering, and not to fusion. In this way, the absorption cross section
evaluated with $W_{\rm surf}(R)$ is related to the total reaction cross section,
rather than CF.\\

\section{Conclusions}

We investigated nucleus-nucleus potentials for approximate calculations of fusion and elastic scattering.\\

In coupled-channel calculations taking into account all relevant direct reaction channels, one can use a
bare potential of a systematic approach, like the folding model, together with a strong imaginary potential
with a short range. Then, the resulting cross sections are expected to give reasonable descriptions of the
data. However, some relevant direct reaction channels are frequently left out of the model space. In such
cases, it may be necessary to account for the influence of these channels in an approximate way, 
and different procedures are required for elastic scattering and fusion cross sections.\\

The elastic scattering cross section is fully determined by the radial wave functions in the elastic channel.
Thus, the effects of all relevant channel couplings must be taken into account. Usually, the influence 
of channels neglected in the model space is simulated by the addition of a complex polarization potential 
to the bare interaction, or by replacing it with a phenomenological optical potential. Both procedures extend the 
range of the imaginary potential, reaching the barrier region and beyond. Thus, absorption simulates both 
fusion and direct reactions to
neglected channels.\\

In the case of fusion, the situation is quite different. The fusion process can take place in the entrance
channel and also after the transition to an inelastic channel. Thus, the fusion cross section depends on
the wave functions of all channels in the model space. Therefore, neglected channels cannot be replaced by
a complex polarization potential. We have shown that this replacement is much worse than completely ignoring 
the influence of the couplings. Above the Coulomb barrier, couplings with bound channels do not lead
to appreciable changes in the cross section. However, it may be very important at sub-barrier
energies. We have shown that excellent results can be obtained by taking into account the real part
of the polarization potential but ignoring its imaginary part.

\section*{Acknowledgments}

Work supported in part by the Brazilian funding agencies, CNPq, FAPERJ, CAPES, and the INCT-FNA (Instituto Nacional de Ci\^encia e 
Tecnologia-F\'\i sica Nuclear e Aplica\c c\~oes), research project 464898/2014-5.
The work of F.M.N. was in part supported by the U.S. Department of Energy grant DE-SC0021422 and National Science Foundation CSSI program under award No. OAC-2004601 (BAND Collaboration). 


\bibliographystyle{apsrev}
\bibliography{fusbreak-2025_vs05} 
\end{document}